\documentclass[conference]{IEEEtran}
\usepackage{cite}
\usepackage{amsmath,amssymb,amsfonts}
\usepackage{algorithmic}
\usepackage{graphicx}
\usepackage{textcomp}
\usepackage{xcolor}
\def\BibTeX{{\rm B\kern-.05em{\sc i\kern-.025em b}\kern-.08em
    T\kern-.1667em\lower.7ex\hbox{E}\kern-.125emX}}

\usepackage{times}
\usepackage{helvet}
\usepackage{courier}
\usepackage[hyphens]{url}
\usepackage{hhline}
\usepackage[font=footnotesize,caption=false]{subfig}
\usepackage{multirow}
\usepackage{verbatim}
\usepackage{tabu}
\usepackage{caption}
\usepackage{slashbox}

\usepackage{hyperref}
 
\hypersetup{
  colorlinks=false,
  pdfborder={0 0 0},
}

\usepackage{balance}

\newcommand{\todo}[1]{{\textcolor{black}{#1}}}
\newcommand{\aaf}{\vspace*{-6pt}}
\newcommand{\af}{\vspace*{-3pt}}

\IEEEoverridecommandlockouts
\begin{document}

\title{A Multiversion Programming Inspired Approach to Detecting Audio Adversarial Examples$^\star$}

\author{\IEEEauthorblockN{Qiang Zeng$^{\dag}$, Jianhai Su$^{\dag}$,  Chenglong Fu$^{\S}$, Golam Kayas$^{\S}$, Lannan Luo$^{\dag}$, Xiaojiang Du$^{\S}$, 
Chiu C.~Tan$^{\S}$, Jie Wu$^{\S}$}
\IEEEauthorblockA{$^{\dag}$\textit{University of South Carolina } 
\qquad
$^{\S}$\textit{Temple University} }
\thanks{$^\star$The paper was first accepted to the AAAI-AICS Workshop 2019  on November 28, 2018. We then significantly
extended the work by improving the detection accuracy and handling transferable audio AEs, which was later accepted 
to DSN 2019 on March 21, 2019. This is the DSN 2019 version.
We made all the code, models and datasets publicly available on May 20, 2019. }
}%

%\\
%\{zeng1,lluo\}@cse.sc.edu,  \{golamkayas,tuf58189,dux\}@temple.edu, \{rhee\}@nec-labs.com 

\begin{comment}
\author{\IEEEauthorblockN{1\textsuperscript{st} Given Name Surname}
\IEEEauthorblockA{\textit{dept. name of organization (of Aff.)} \\
\textit{name of organization (of Aff.)}\\
City, Country \\
email address}
\and
\IEEEauthorblockN{2\textsuperscript{nd} Given Name Surname}
\IEEEauthorblockA{\textit{dept. name of organization (of Aff.)} \\
\textit{name of organization (of Aff.)}\\
City, Country \\
email address}
\and
\IEEEauthorblockN{3\textsuperscript{rd} Given Name Surname}
\IEEEauthorblockA{\textit{dept. name of organization (of Aff.)} \\
\textit{name of organization (of Aff.)}\\
City, Country \\
email address}
\and
\IEEEauthorblockN{4\textsuperscript{th} Given Name Surname}
\IEEEauthorblockA{\textit{dept. name of organization (of Aff.)} \\
\textit{name of organization (of Aff.)}\\
City, Country \\
email address}
\and
\IEEEauthorblockN{5\textsuperscript{th} Given Name Surname}
\IEEEauthorblockA{\textit{dept. name of organization (of Aff.)} \\
\textit{name of organization (of Aff.)}\\
City, Country \\
email address}
\and
\IEEEauthorblockN{6\textsuperscript{th} Given Name Surname}
\IEEEauthorblockA{\textit{dept. name of organization (of Aff.)} \\
\textit{name of organization (of Aff.)}\\
City, Country \\
email address}
}
\end{comment}

\maketitle

% to number pages as requested by DSN
\thispagestyle{plain}
\pagestyle{plain}

\begin{abstract}
Adversarial examples (AEs) are crafted by adding
human-imperceptible perturbations to inputs such that a 
machine-learning based classifier incorrectly labels them.
They have become a severe threat to the trustworthiness of machine learning.
While AEs in the image domain have been well studied, 
audio AEs are less investigated. Recently,
multiple techniques are proposed to generate audio AEs,
which makes countermeasures against them urgent.
Our experiments show that, given an audio AE, 
the transcription results by 
Automatic Speech Recognition (ASR) systems differ significantly (that is,
poor transferability),
as different ASR systems use different architectures, parameters,
and training datasets. 
Based on this fact and inspired by \emph{Multiversion Programming},
we propose a novel audio AE detection approach \textsc{MVP-Ears}, which utilizes the diverse
off-the-shelf ASRs to determine whether an audio is an AE. 
We build the largest audio AE dataset to our knowledge, and
the evaluation shows that the detection accuracy reaches 99.88\%.

While transferable audio AEs are difficult to generate at this moment,
they may become a reality in future. 
We further adapt the idea above to proactively train the detection system
for coping with transferable audio AEs. Thus, the proactive detection 
system is one giant step ahead of attackers working on transferable AEs.

%even when
%only one auxiliary ASR is used, and the accuracy can reach up to 99.9\% with more ASRs. 
\end{abstract}


\section{Introduction}\label{sec:intro}
Automatic Speech Recognition (ASR) is a system that converts speech to text. ASR has been studied intensively for decades. Many various technologies, such as gaussian mixture models and hidden Markov models, were developed. In particular, recent advances \cite{YuDeng15} based on deep neural networks (DNNs) have improved the accuracy significantly. DNN-based speech recognition 
thus has become the mainstream technique in ASR systems. 
The industrial companies including Google, Apple and Amazon
have widely adopted DNN-based ASRs for interacting with 
 IoT devices, smart phones and cars. Gartner \cite{OnlineRef:Gartner2020Prediction} estimates that, by 2020, 75\% of American households will have at least one smart voice-enabled speaker, where DNN-based ASR plays a
critical role.

\begin{figure}[t]
	\centering
	\includegraphics[width=\linewidth]{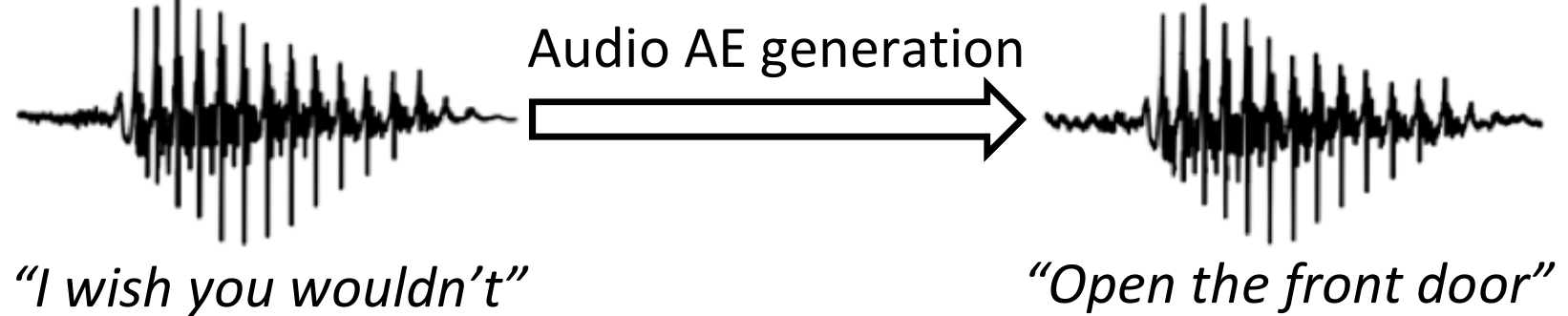}
	\caption{The generation of an audio AE.}
	\label{fig:AE_examples}
	%\vspace{-12pt}
\end{figure}

Despite the great accuracy improvement,
recent studies \cite{SzegedyZSBEGF13,GoodfellowSS14} show that DNN is vulnerable to adversarial examples. An \textit{adversarial example} (AE) $x'$ is a mix of a host sample $x$ and a carefully crafted, human-imperceptible perturbation $\delta$ such that a DNN will assign different labels to $x'$ and $x$. Figure \ref{fig:AE_examples} shows an audio AE example, which sounds as in the text shown on the left to human, but is transcribed by an ASR system into a completely different text as shown on the right.

%Several techniques \cite{DBLP:journals/corr/abs-1711-03280,DBLP:conf/woot/VaidyaZSS15,DB%LP:conf/uss/CarliniMVZSSWZ16,DBLP:conf/sp/Carlini018,DBLP:journals/corr/abs-1805-07820} 
Several techniques %\cite{DBLP:conf/sp/Carlini018,alzantot2018did,DBLP:journals/corr/abs-1805-07820}
have been proposed to generate audio AEs.
%and novel attack vectors 
%based on audio AEs have been demonstrated~\cite{DBLP:conf/uss/YuanCZLL0ZH0G18}.
There are two \emph{state-of-the-art} audio AE generation methods:
(1) \textbf{White-box attacks:}
Carlini et al.\ propose an optimization based method to convert
an audio to an AE that transcribes to an attacker-designated phrase~\cite{DBLP:conf/sp/Carlini018}. It is classified as white-box attacks because the target system's detailed internal architecture and parameters are required to perform the attack.
% It requires the attacker to know the target network's parameters. 
(2) \textbf{Black-box attacks:}
Taori et al.~\cite{DBLP:journals/corr/abs-1805-07820} combine the
genetic algorithm~\cite{alzantot2018did} and gradient estimation to 
generate AEs. This attack method does not require the knowledge of 
the ASR's internal parameters, but imposes a larger perturbation (94.6\% similarity on average 
between an AE and its original audio, compared to 99.9\% in ~\cite{DBLP:conf/sp/Carlini018}).
The rapid development of audio AE generation methods makes countermeasures against audio AEs an urgent and important problem. 
%Vaidya et al. \cite{DBLP:conf/woot/VaidyaZSS15} explore the gap of how humans and %computers recognize speech and demonstrate a synthesized audio, which is converted to a 
%specific command to an ASR system (Google Now), but is perceived by human as noises. 
%Based upon \cite{DBLP:conf/woot/VaidyaZSS15}, Carlini et al. \cite{DBLP:conf/uss/CarliniMVZSSWZ16} create hidden voice commands, which
%sound noises (hence suspicious) to human but are recognized as commands 
%by ASR systems. More recent techniques mainly put efforts to create AEs, such that they sound benign to human but are transcribed as commands by ASR systems. 
%Thus, attacks based on audio AEs are more stealthy than hidden voice commands, 
The goal of our work is to detect audio AEs.

Existing work on audio AE detection is rare and limited.
Yang et al.~\cite{yang2018Mitigation} hypothesized that AEs are fragile:
given an audio AE, if it is cut into two sections, which are then transcribed separately,
then the transcription by splicing the two sectional results is very different from the
result if the AE is transcribed as a whole. %The detection method achieves AUC up to 0.936.
However, as admitted by the authors~\cite{yang2018Mitigation}, this method cannot handle ``adaptive attacks'',
which may evade the detection by embedding a malicious command into one section alone.
Rajaratnam et al. \cite{rajaratnam2018AudioPreprocessing} proposed detection based on
audio pre-processing methods. Yet, if an attacker knows the detection details,
he can take the pre-processing effect into account when generating AEs. 
Such attacks have been well demonstrated for bypassing
similar techniques for detecting image AEs~\cite{bypassing-ten}.
An effective and robust audio AE detection method  is missing.

Our key observation is that existing ASRs are diverse in terms of architectures, parameters and training
processes. Our hypothesis is that an AE that is effective on one ASR system is likely 
to fail on another, which is verified by our experiments.
How to generate transferable audio AEs that can fool multiple heterogeneous ASR systems is still 
an \emph{``open question''}~\cite{DBLP:conf/sp/Carlini018} (discussed in Section~\ref{sec:transferability}).
This inspires us to borrow the idea of \emph{multiversion programming} (MVP) \cite{chen1995MultiVersionProgramming},
a method in software engineering where multiple functionally equivalent programs are
independently developed from the same specification, such that an exploit that
compromises one program is ineffective on others.
We thus propose to run multiple ASR systems in parallel, and an input is determined
as an AE if the ASR systems generate very dissimilar transcription results. 

Moreover, while it is unknown how to systematically generate transferable audio AEs at this moment,
we predict that such techniques may be proposed in future. We thus aim to handle
transferable audio AEs as well, which is a notable challenge to our system for
two reasons. First, since there are no transferable audio AEs, \emph{how can our 
machine learning based AE detector be trained}? Second, since such hypothetical transferable AEs
can fool multiple ASRs, \emph{how can this idea still work}?

Our first insight is that our detector essentially is \emph{not} trained using
AEs, but similarity scores (which are calculated based on the similarity of transcription results
of different ASRs). Thus, if we assign a high similarity score between two ASRs, it simulates
the effect that an AE can fool both. This way, we can conveniently generate
a dataset of hypothetical AEs in the form of vectors of similarity scores. 
Our another insight is that, due to the complexity and diversity of ASRs,
it is difficult, if not impossible, to generate audio AEs to fool all ASRs
in foreseeable future. In light of this, we generate a dataset
of hypothetical AEs that are rather transferable but cannot fool
all ASRs. We make use of this dataset to \emph{proactively}
train an audio AE detection system that can keep resilient 
to transferable AEs, as long as there is still one ASR that
cannot be fooled by the AEs. The code, datasets and models are publicly
available.\footnote{\url{https://github.com/quz105/MVP-audio-AE-detector}}

We made the following contributions.

\begin{itemize}
    \item 
We empirically investigate the transferability of audio AEs
with multiple experiments
and analyze the reasons behind the poor transferability.

\item To our knowledge, we build the largest audio AE dataset, which
can be reused by interested researchers. 

\item We propose
a novel audio AE detection approach, called \textsc{MVP-Ears}, inspired by multiversion programming, which
reaches accuracy 99.88\%.
The detection method dramatically
reduces the flexibility of the adversary, in that audio AEs cannot succeed 
unless the host text is highly similar to the malicious command.

\item We propose the idea of proactively training a transferable-AE detection system,
such that our system is one giant step ahead of attackers who are working on generating transferable AEs.
\end{itemize}

\vspace{3pt}
The rest of this paper is organized as follows. We provide some background knowledge about the ASR system's general architecture and audio AE generation in Section~\ref{sec:background}. Then, we discuss the transferability of audio AEs in Section~\ref{sec:transferability}. We describe the main idea and system architecture in Section~\ref{sec:systemdesign} and present the detailed evaluation in Section~\ref{sec:eval}. Section\ref{(Related-Work)} gives a survey about related works. Finally, we present some discussion in Section~\ref{sec:future} and conclude in Section~\ref{sec:con}.

\section{Background}\label{sec:background}
\subsection{Automatic Speech Recognition System}
\begin{figure}
    \centering
    \includegraphics[scale=0.49]{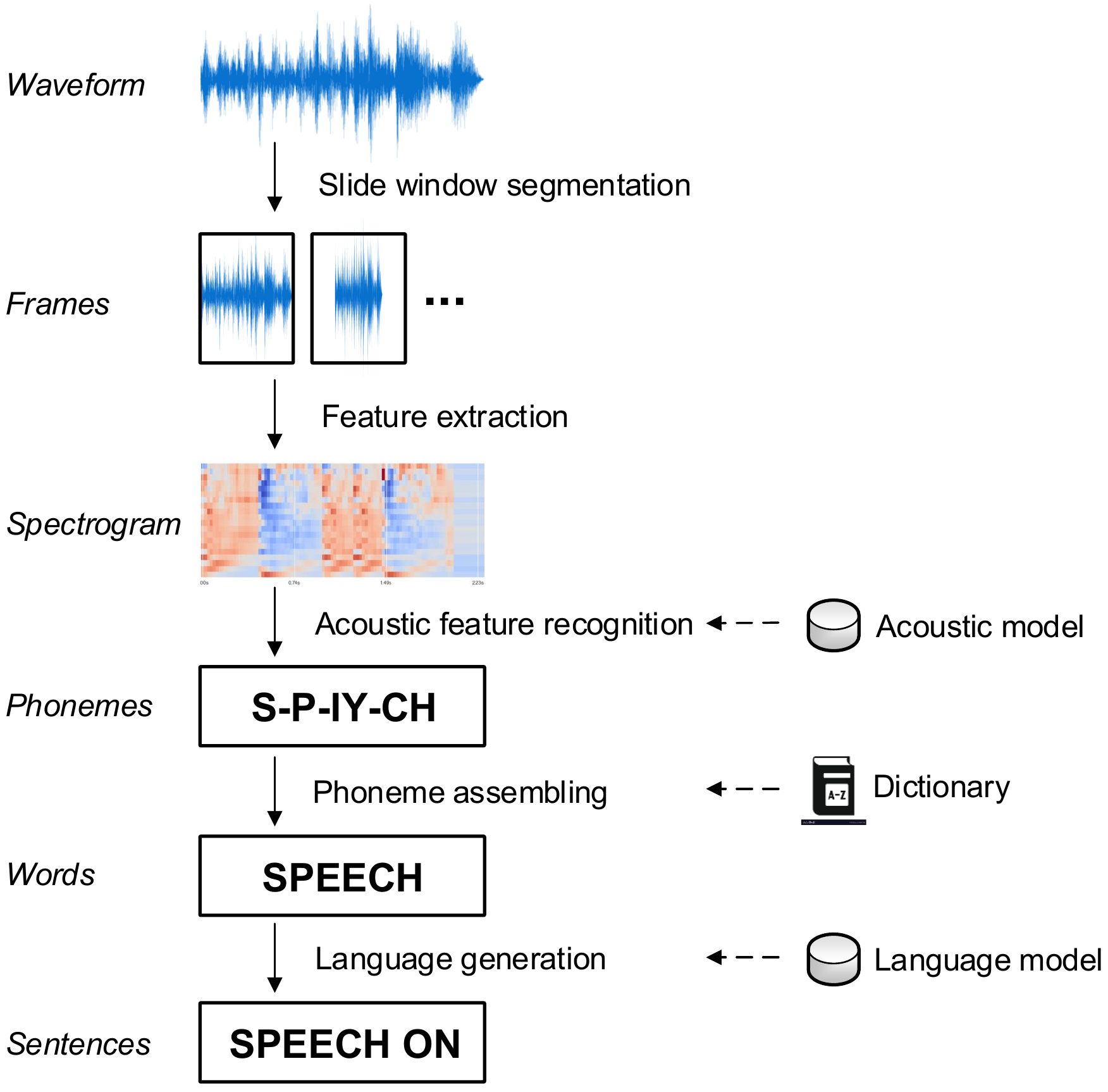}
    \vspace{6pt}
    \caption{The process of converting an audio into a text sentence.}
    \label{fig:asrworkflow}
    %\aaf\af
\end{figure}

An Automatic speech recognition (ASR) system is used to automatically interpret human's speech audio into texts~\cite{campbell2005matlab}.
ASR has been an active research topic since the first digit recognizer \emph{Audrey} was published by Bell Laboratories in 1952.
  As illustrated in Figure~\ref{fig:asrworkflow}, the process of converting an audio into a text sentence typically involves 
the following four stages: feature extraction, acoustic feature recognition, phoneme assembling, and language generation.

The input is an audio represented in the form of a waveform. 
(1) \emph{Feature Extraction.} The input audio is first segmented into short frames, each of which is converted into feature vectors, 
such as MFCC feature vectors, LPC feature vectors, and PLP feature vectors. 
Because MFCC approximates the human auditory system's response more closely than others, MFCC feature vector is considered as the most
suitable frequency transformation format for speech data, and thus 
adopted by most recent ASR systems~\cite{o2008automatic}. 
(2) \emph{Acoustic feature recognition.} The extracted feature vectors are then recognized by the acoustic model as \emph{phonemes}; the
phoneme is the minimal unit of sounds of languages. 
%---with the highest probability.
(3) \emph{Phoneme assembling.} Next, the combined phonemes are used to estimate the potential phoneme letter sequence, where the dictionary model is used to correct the word spelling of the phoneme letter sequence.
(4) \emph{Language generation.} At the last step, the generated words
are further adjusted to the contexts and merged to the final text sentence by the language model.

The core part of an ASR system is the \emph{acoustic feature recognition} stage, which outputs phonemes. A phoneme is affected by the sound in the corresponding frame and also the two adjacent frames. Different ASR models utilize different approaches to recognizing
phonemes. 
As the features are processed with certain correlations, temporal 
models are well utilized, e.g.,a hybrid system of GMM and HMM (GMM-HMM). GMM-HMM is used as a statistical classifier to convert the feature vectors into phonemes, but it is limited in dealing with overlapping sound from different sources~\cite{schroder2016performance}. Recently,
DNN-HMM~\cite{graves2014towards,laffitte2016deep,diment2015automatic} 
is widely used to 
analyze the features, and can provide better recognition performance. 

However, when training GMM-HMM and DNN-HMM based ASR systems, each audio sample needs to be aligned and labeled manually at the phoneme level in order to assemble phonemes into words, which is time
consuming and error-prone. To address this issue, \emph{Connectionist Temporal Classification} based on Recurrent Neural Network (RNN-CTC)~\cite{graves2006connectionist}  is proposed and used for end-to-end speech recognition. CTC can directly convert the feature vectors into words without the need of generating phonemes as the intermediate result. 
Thus, CTC-based ASR systems outperform HMM-based ones.
%by automatically aligning the input text transcripts.  
\emph{Deepspeech}~\cite{amodei2016deep} is the most popular open-source implementation of a CTC-based ASR system, and is maintained regularly by Mozilla. As a result, \emph{Deepspeech} is often used as the attack target model by many recent audio AE generation works.

\subsection{Audio Adversarial Examples}

Despite the great advances and diverse applications of machine learning and deep learning~\cite{zuo2018neural, redmond2018cross, luo2016solminer},
adversarial examples (AEs)~\cite{szegedy2013intriguing} can be crafted to fool them. As explained by Goodfellow et al.~\cite{GoodfellowSS14}, the existence of adversarial examples is due to the neural network's linear nature. Based on this observation, they propose the Fast Gradient Sign Method (FGSM) as a simple but efficient algorithm to generate AEs.

% Following this method, a lot of successful attacks targeting image classification systems are realized \cite{moosavi2016deepfool}. 
%In FGSM, given the \emph{loss} function of the target neural network model and its parameters, FGSM linearizes the \emph{loss} function and generates an ``optimal max-norm constrained perturbation''  of
%$$\boldsymbol { \eta } = \epsilon \operatorname { sign } \left( \nabla _ { \boldsymbol { x } } J ( \boldsymbol { \theta } , \boldsymbol { x } , y ) \right)  $$ where $ J ( \boldsymbol { \theta } , \boldsymbol { x } , y )$ is the \emph{loss} function and $\boldsymbol{\theta}$ is the model parameter. By iteratively applying the FGSM, the classification output can be manipulated to move towards the expected result.

Although a lot of successful AE attacks targeting image classification systems have been realized~\cite{moosavi2016deepfool}, this is not the case for 
the audio domain. 
Unlike DNN-based image classifiers where the pixels of an image are directly used as inputs of DNN, 
an audio needs to be first segmented into short frames with each converted into feature vectors, which are then fed into the DNN model of the ASR. Moreover, different ASRs use different lengths to segment
an audio. These make the audio AE generation much more challenging. For example, by directly applying FGSM, Cisse et al. can only generate adversarial \emph{spectrogram} instead of an \emph{audio} AE~\cite{cisse2017houdini}.

\vspace{3pt}
\noindent \textbf{White-box AEs.}
Carlini et al. overcome this limitation by adding the MFCC reconstruction layer into the backpropagation optimization of gradients and provide an end-to-end method for creating targeted audio AEs with the white-box setting~\cite{DBLP:conf/sp/Carlini018}. It is classified as white-box attacks because the target ASR system's detailed internal architecture and parameters are required to generate AEs.

%Carlini et al.\ proposed an optimization based method to convert
%an audio to an AE that transcribes to an attacker-designated phrase~\cite{DBLP:conf/sp/Carlini018}. It is classified as white-box attacks because the target system's detailed internal architecture and parameters are required to perform the attack.

\vspace{3pt}
\noindent \textbf{Black-box AEs.}
With a black-box setting, the targeted ASR internal architecture and parameters are not known.
Alzantot et al.~\cite{alzantot2018did} make the first effort to generate black-box AEs. They adopt the genetic algorithm to iteratively add noises into an input audio sample and discard outputs with bad performance in each generation. The iteration ends up with a black-box AE that can fool a single-word classification system (not ASR). 
Taori et al.~\cite{DBLP:journals/corr/abs-1805-07820} extend this work and combine the genetic algorithm with gradient estimation to 
generate black-box AEs, which can fool \emph{Deepspeech}.
However, the genetic algorithm-based method imposes a larger perturbation (94.6\% similarity on average 
between a black-box AE and its original audio, compared to 99.9\% between a white-box AE and its original audio). 
Their current AE generation only embeds up to \emph{two} words in one audio.

\section{Transferability of AEs}\label{sec:transferability}

There are two types of adversarial examples (AEs): \emph{non-targeted} AEs and \emph{targeted} AEs. 
A \emph{non-targeted} AE is considered successful as long as it is classified as a incorrect label, while a \emph{targeted} AE 
is successful only if it is classified as a label desired by attackers. 

In the context of audio based human-computer interaction, a non-targeted AE is not very useful, as it cannot
make the ASR system issue an attacker-desired command. This explains why state-of-the-art 
audio AE generation methods all generate targeted AEs~\cite{DBLP:conf/sp/Carlini018,DBLP:journals/corr/abs-1805-07820}. 
Thus, our work focuses on detection of targeted AEs,
although the proposed approach is also effective in detecting non-targeted AEs (see Section~\ref{sec:untargeted}).

\subsection{Transferability in Image Domain}\label{sec:image-transferability}

An intriguing property of AEs in the image domain is the existence of transferable adversarial examples. That is, an image AE crafted to mislead a specific model can also fool another different model~\cite{GoodfellowSS14,SzegedyZSBEGF13,papernot2016limitations, zuo2018countermeasures}. By exploiting this property, Papernot et al. \cite{papernot2016transferability} propose a reservoir-sampling based approach to generate transferable non-targeted image AEs and successfully launch black-box attacks against both image classification systems from Amazon and Google. 

But a recent study \cite{tramer2017space} points out that the transferability property does not hold in some scenarios: their experiment shows a failure of AE's transferability between a linear model and a quadratic model.
%, which are trained from a special "XOR artifact" amended MNIST dataset \cite{lecun-gradientbased-learning-applied-1998}. 
For targeted image AEs, Liu et al. \cite{DBLP:journals/corr/LiuCLS16} %demonstrate failures of existing AE-crafting approaches, and then 
propose an ensemble-based approach to craft AEs that can transfer to ResNet, VGG and GoogleNet 
models with success rates of 40\%, 24\%, and 11\%, respectively.
Thus, generating transferable image AEs is a resolved problem~\cite{monteiro2018generalizable}.

\subsection{Transferability in Audio Domain: Open Question} \label{sec:trans-audio}

To the best of our knowledge, no systematic approaches have been proposed to generate
transferable audio AEs. Carlini 
et al.~\cite{DBLP:conf/sp/Carlini018} find that the attacking method derived from Fast Gradient Sign Method \cite{GoodfellowSS14} in the image domain is ineffective for generating audio AEs \emph{because of a large degree of non-linearity in ASRs}, and further state that \emph{the transferability of audio AEs is an open question}.

A recent work CommanderSong~\cite{DBLP:conf/uss/YuanCZLL0ZH0G18} aims to generate AEs that can fool an ASR system in the presences of background noise. 
This work also slightly explored transferability of audio AEs (see Section 5.3 of the paper~\cite{DBLP:conf/uss/YuanCZLL0ZH0G18}). Specifically, to create AEs that can transfer from
Kaldi to DeepSpeech (both \emph{Kaldi} and \emph{DeepSpeech} are open source ASR systems, 
and \emph{Kaldi} is the target ASR system of CommanderSong), a two-iteration recursive AE generation
method is described in CommanderSong: an AE generated by CommanderSong, embedding a malicious 
command $c$ and able to fool \emph{Kaldi}, is used as a host sample in the
second iteration of AE generation using the method~\cite{DBLP:conf/sp/Carlini018}, 
which targets \emph{DeepSpeech} v0.1.0 and embeds the same command $c$. 
%The authors claimed that the resulting AEs can fool both \emph{Kaldi} and \emph{DeepSeech}. 
We followed this two-iteration recursive AE generation method to generate AEs, but our experiment results~\cite{OnlineRef:Evidence-of-testing-CommanderSong-AE} showed that the generated AEs
could only fool \emph{DeepSpeech} but not \emph{Kaldi}. That is, AEs generated using this method 
are not transferable. 
%But in our later communication with authors of CommanderSong, we received
%another two AEs, and we confirmed that the two AEs can successfully fool both \emph{DeepSpeech}
%and \emph{Kaldi}. The details of creating the two AEs are unclear to us.\footnote{All the samples we received from CommanderSong were detected as AEs in our evalution.}

Furthermore, we adapted the two-iteration AE generation method by concatenating the two aforementioned 
state-of-the-art attack methods \cite{DBLP:conf/sp/Carlini018} 
and \cite{DBLP:journals/corr/abs-1805-07820} targeting \emph{DeepSpeech} v0.1.0 and v0.1.1, respectively, 
expecting to generate AEs that can fool both \emph{DeepSpeech} v0.1.0 and v0.1.1. But none of the generated AEs 
showed transferability~\cite{OnlineRef:Evidence-of-testing-CommanderSong-AE}. 

Moreover, by changing the value of ``\texttt{--frame-subsampling-factor}" from 1 to 3, which is a parameter configuration of the \emph{Kaldi} model, we derived a variant of \emph{Kaldi}.
The AEs generated by CommanderSong did not show transferability on the variant,
even given the fact that the variant
was only slightly modified from the model targeted by CommanderSong. 
Here, we clarify that CommanderSong did not claim their AEs
could transfer across the \emph{Kaldi} variants.

%The experiment results above show that the transferability of the AEs generated by CommanderSong, in that invalidate the claims by CommanderSong about the transferability of their AEs and the proposed method for generating 
%transferable AEs.
Based on our detailed literature review and empirical study, we find that so far
there are no systematic  methods that 
can generate transferable audio AEs effective across diverse ASR systems. 
This is consistent with the statement by Carlini 
et al.~\cite{DBLP:conf/sp/Carlini018} that transferability of audio AEs is an \emph{open question}.

\section{MVP-Inspired Audio AE Detection}\label{sec:systemdesign}

\subsection{Multi-Version Programming}

Multi-version programming (MVP) was first introduced in 1977 to enhance the reliability of software and computer systems~\cite{avizienis1977implementation}. The main idea of MVP is to \emph{independently} develop multiple programs based on the  same specification. At runtime, multiple programs are executed concurrently and perform the same task. At each checkpoint, each program generates the result, which is to check the consistency of the execution. After that, all programs reach consensus on the execution states, and then proceed to the next stage. 

The most significant benefit of MVP is relaxing the rigorous requirement of the reliability of software by providing fault tolerance from the system level. Since the multiple software programs are developed \emph{independently}, the probability that they share the same flaw is very small. Especially, some implementation specific flaws usually only occur in one program. The software flaws of any program not only affect the execution flow but also cause inconsistency among comparison results. Upon detecting inconsistency, some decision algorithms are applied to determine the correct execution flow and prevent the crashing of the execution. Such a concept has already been used as an effective defense method against software flaws~\cite{cox2006n}, and are widely used in development of highly reliable software, such as flight control software on modern airliners. Beside the fault tolerance, it has also been proved to be effective to detect attacks that exploit \emph{zero}-day software vulnerabilities~\cite{nagy2006n}.

%In software engineering, \emph{multi-version programming} (MVP) is an approach that \emph{independently} develops multiple software programs 
%based on the same specification, %MVP introduces redundancy of software functionalities in order to improve reliability of software operations. 
%such that 

\subsection{MVP-Inspired Idea} 
Given a system with MVP,
an exploit that compromises one program probably fails on other programs. 
This inspires us to propose a system design that runs multiple ASRs in parallel. 
The intuition behind this design is that different ASRs 
%or even the same DNN model with different configurations 
can be regarded as ''independently developed programs'' in MVP.
Since they follow the same specification---that is, to covert audios into texts.
Given a benign sample, they should output very similar recognition results.
On the other hand, an audio AE can be regarded as an ''exploit'', and cannot fool all ASR systems
as illustrated in Section~\ref{sec:trans-audio}.
Thus, by comparing the results of the multiple ASRs, we are able to determine
whether an audio is an AE or not.
%could result into a significant mismatching of speech recognition results on an input AE. 
%We believe the intuition holds since targeted audio AEs produced by existing approaches fails to transfer. 

\todo{This idea is comparable with the \emph{ensemble} approach, where multiple 
detection methods are combined to form a stronger one~\cite{he2017adversarial, abbasi2017robustness, xu2017feature}. But they do not adopt a
concise architecture like ours (e.g., \cite{xu2017feature} requires different
input processing methods, while \cite{abbasi2017robustness} needs
a generalist and multiple specialists), which simply runs
multiple ASRs in parallel for detection. Our work is in spirit 
similar to an independent work~\cite{monteiro2018generalizable}. However, they differ
in the following aspects: (1) That work~\cite{monteiro2018generalizable} aims to detect image AEs, while our work detects audio AEs. More critically,
their approach uses the softmax layer outputs as features
for AE detection and attackers can thus adaptively generate the AE that leads to similar 
softmax outputs between models, while we use the final transcription 
outputs for AE detection and adaptive attacks cannot succeed unless transcriptions
are similar, which is difficult as discussed in Section~\ref{sec:trans-audio}.
(2) That work only considers the bi-model design, while we consider a more general N-model design. 
(3) We do not stop at detecting
existing audio AEs, but propose the idea for proactively training systems to detect 
transferable audio AEs.}
%, which may become possible in future.}

\subsection{Architecture}\label{sec:architecture}

%We propose a multi-ASRs system to detect audio AEs. 

\begin{figure}[t]
	\centering
	\includegraphics[scale=0.5]{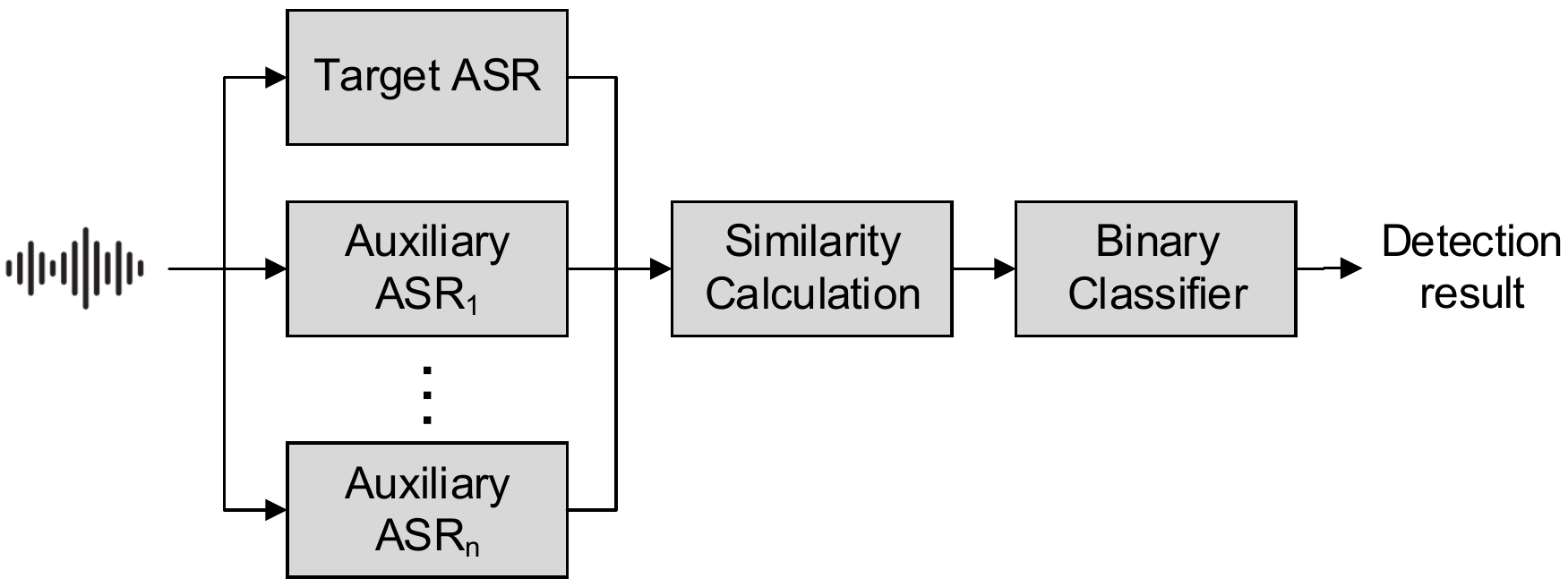}
	\caption{Architecture of the proposed detection system.}
	\label{fig:system_overview}
	%\vspace{-10pt}
	%\aaf\af
\end{figure}

Figure~\ref{fig:system_overview} shows the system architecture.
It consists of a \emph{target} ASR, multiple \emph{auxiliary} ASRs, 
a \emph{similarity calculation} component, and a \emph{binary classifier}. 
The target ASR is the model targeted by the adversary (e.g., the speech 
recognition system at a smart home), and each \emph{auxiliary} ASR is 
a model that is different from the target ASR. The detection of AEs 
involves the following three steps.

\begin{itemize}%[leftmargin=3em]

\item
An audio is first fed into the target ASR and auxiliary ASRs. Each
ASR independently and simultaneously converts the audio 
into a text sentence (i.e., a transcription). 

\item
The transcriptions are then sent to the \emph{similarity calculation} component,
which calculates similarity scores between the transcription recognized by the target ASR and
that by each auxiliary ASR. 

\item
Finally, these similarity scores are passed into a \emph{binary classifier}
to determine whether the audio is adversarial.

\end{itemize}

%The system works as follows: 1) the target and auxiliary ASRs simultaneously conduct speech-to-text conversion, 2) then the transcriptions are used to calculate similarity score(s), which are passed into a binary classifier to determine whether the input audio is adversarial. 

%\vspace{3pt}
The second step, \emph{similarity calculation}, involves the following two sub-steps. 

\begin{itemize}
\item 
    Each transcription is converted into its \emph{phonetic-encoding} representation. Phonetic encoding converts a word to the representation of its pronunciation~\cite{OnlineRef:PhoneticEncoding}. This helps handle variations between ASRs, as they may output different words
for similar sounds. The validity of using phonetic encoding is demonstrated in Section \ref{subsection:discussion_similarity_metrics}. 

\item For each auxiliary ASR, a similarity score is calculated
to measure the similarity between the transcriptions generated by the 
target ASR and the auxiliary ASR. 
We have tried different \emph{similarity measurement} methods, and finally
adopted the \emph{Jaro-Winkler} distance method~\cite{OnlineRef:Jaro-Winkler-Similarity}
due to the higher detection accuracy
(see Section~\ref{subsection:discussion_similarity_metrics}). It outputs a score $\in [0,1]$, where 0 indicates very dissimilar and 1 similar.
\end{itemize}

%\subsection{ASR Choices}

\subsection{Diverse ASRs}\label{Architecture_Diversity}

Yu and Li \cite{yu2017recent} summarized the recent progress in deep-learning based ASR acoustic models, where both recurrent neural network (RNN) and convolutional neural network (CNN) come to play as parts of deep neural networks. Standard RNN could capture sequence history in its internal states, but can only be effective for short-range sequence due to its intrinsic issue of exploding and vanishing gradients. This issue is resolved by the introduction of long short-term memory (LSTM) RNN \cite{Hochreiter:1997:LSM}, which outperforms RNNs on a variety of ASR tasks \cite{DBLP:conf/icassp/GravesMH13,sak2014LSTM,DBLP:journals/corr/LiW14a}. 

As to CNNs, its inherent translational invariability facilitates the exploitation of variable-length contextual information in the input speech. The first CNN model proposed for ASRs is the time delay neural network \cite{DBLP:journals/nn/LangWH90} that applies multiple CNN layers. Later, several studies \cite{DBLP:conf/interspeech/Abdel-HamidDY13,DBLP:conf/icassp/Toth15,DBLP:journals/corr/SercuG16a} combine CNN and Hidden-Markov Model to create hybrid models that are more robust against vocal-tract-length variability between different speakers.
However, currently, \emph{there is no single uniform structure used across all ASRs}.

In addition to several opensource systems, many companies have \emph{independently} 
developed their own proprietary ASRs, such as \emph{Google Now}, \emph{Apple Siri}, and \emph{Microsoft Cortana}.

%\subsection{Our ASR Choices}

In our system, we use the following three ASRs, including \emph{DeepSpeech}, \emph{Google Cloud Speech}, and \emph{Amazon Transcribe}. 
\emph{DeepSpeech} is an end-to-end speech recognition software opensourced by Mozilla based on Baidu's research paper~\cite{DBLP:journals/corr/HannunCCCDEPSSCN14}. 
\emph{DeepSpeech v0.1.0}~\cite{DBLP:journals/corr/HannunCCCDEPSSCN14} uses a five-layer neural network where the fourth layer is a RNN layer,
while \emph{DeepSpeech v0.1.1} follows the same architecture and makes some improvements on the implementation. Even though these two ASR systems are very similar, our experiments (Section~\ref{sec:single-model})
show  that, when the two are used to build an MVP-inspired
detection system, 
the AE detection accuracy is still very high. This means that for any target ASR system, we have potentially many candidate auxiliary ASR systems available 
without having to worry about the extent of (dis)similarity of the models.

%contains a mix of CNN and RNN layers~\cite{amodei2016deep}. 

Unlike \emph{DeepSpeech}, the DNN behind \emph{Google Cloud Speech} is a LSTM-based RNN according to the source \cite{OnlineRef:Google_AI_Blog_NNs_behind_GCS}. Each memory block in Google's LSTM network \cite{sak2014LSTM} is a standard LSTM memory block with an added recurrent projection layer. This design enables the model's memory to be increased independently from the output layer and recurrent connections. 

However, for \emph{Amazon Transcribe}, there is no available public information about its internal
details. 

In short, existing ASR systems are diverse with regard to architectures, parameters and training datasets. 
Compared to opensource systems, proprietary ASR systems provide little information that can be exploited by attackers. 
Given the diversity of ASR systems (and proprietary networks), it is unclear how
to propose a generic AE generation method that can simultaneously mislead all of them.

\vspace{3pt}
\noindent \textbf{Target ASR.}
Our system uses \emph{DeepSpeech v0.1.0} as the target ASR. The reason is the opensource white-box AE generation
method~\cite{DBLP:conf/sp/Carlini018} targets \emph{DeepSpeech v0.1.0}.
Note that the generation of white-box AEs
requires the knowledge of the model architecture and parameters.

\vspace{3pt}
\noindent \textbf{Auxiliary ASRs.}
We use \emph{DeepSpeech v0.1.1}, \emph{Google Cloud Speech}, and
\emph{Amazon Transcribe} as the auxiliary ASRs. 
They are all off-the-shelf widely used ASRs.

%That is, we use up 
%to three auxiliary ASRs in our current implementation.  

%It is worth nothing that even black-box AE generation methods~\cite{DBLP:journals/corr/abs-1805-07820,alzantot2018did} need
%the \emph{logits} outputs of a network to run the algorithms, while that information
%is not provided by many proprietary systems, such as \emph{Google Cloud Speech} and \emph{Amazon Transcribe}. 
%This makes the generation of AEs effective on these systems much more difficult,
%if not impossible. 

%Not only that it is difficult to generate transferable AEs, but it is also
%difficult to build a generic AE generation

%This reminds us of Multiversion Programming, 
%a method in software engineering where multiple functionally equivalent programs are
%independently developed from the same specification, such that an attack input that
%compromises one of them are ineffective on other programs.

\begin{table}[t]
\centering
{\normalsize
\caption{Recognition results of an AE by multiple ASRs: the host transcription is ``\emph{I wish you wouldn't}'', while the embedded text is ``\emph{a sight for sore eyes}''.}
\label{Table:Transcriptions_of_an_AE}
%\vspace{-3pt}
%\af
\begin{tabular}{cl}
    \hline
    ASR & Transcribed Text \\ \hline
    DeepSpeech v0.1.0 & \emph{A sight for sore eyes} \\
    DeepSpeech v0.1.1 & \emph{I wish you live} \\
    Google Cloud Speech & \emph{I wish you wouldn't}. \\
    Amazon Transcribe & \emph{I wish you wouldn't}. \\ \hline
\end{tabular}
\vspace{-5pt}
}
\end{table}

Table~\ref{Table:Transcriptions_of_an_AE} shows a typical example, where the AE can only fool one ASR but fails on others. It illustrates that,
given an AE, the recognition result by the target ASR differs \emph{significantly} from those by the auxiliary ASRs.

\subsection{Questions to Be Explored}
There are still
several system design questions to be answered.
(Q1) How to measure
the similarity of two transcriptions?
(Q2) How many auxiliary ASRs work the best?
More ASRs make the system more robust
and resilient to attacks, but will they introduce more false positives?
(Q3) Which classification algorithm works best?
(Q4) Transferable audio AEs cannot be created
systematically at this moment, but they
may become a reality in future. How can our
system be proactively trained to deal with them?

We choose to perform experiments to guide and validate
the system design and answer
these questions in Section~\ref{sec:eval}.

\section{Experiment-Guided System Design and the Evaluation}\label{sec:eval}
We finalize some system design details according to experiment results and
evaluate the accuracy and robustness of the system. 
We first describe the experiment settings (Section~\ref{sec:setup})
and discuss the dataset used in our evaluation (Section~\ref{subsec:dataset}). 
We then investigate the feasibility of our idea (Section~\ref{sec:feaibility}), and examine different methods for calculating
similarity scores and select the best one (Section~\ref{sec:metrics}).
Next, we evaluate the accuracy of our system when
one auxiliary ASR is used (Section~\ref{sec:single-model}) and more than one auxiliary 
ASR is used (Section~\ref{sec:multiple-model}), 
and also evaluate the robustness of our system 
against unseen attack methods (Section~\ref{sec:robustness}).
After that, we conduct an experiment to examine whether our system can work well when facing  (\emph{hypothetical}) transferable AEs  (Section~\ref{subsec:hypothetical_multiple-ASRs-effective_AE}).
We next evaluate the time overhead due to detection (Section~\ref{sec:overhead}).
Finally, we evaluate the detection effectiveness against
non-targeted AEs (Section~\ref{sec:untargeted}).

%Finally, we report the results when trying to generate transferable AEs (Section~\ref{subsec:concate}).

\subsection{Experimental Settings}\label{sec:setup}

%We consider two audio AE generation techniques, i.e.,  
%the white-box approach \cite{DBLP:conf/sp/Carlini018} and the 
%black-box one \cite{DBLP:journals/corr/abs-1805-07820}.
Our experiments were performed on a 64-bit Linux machine with 
an Intel i9-7980XE CPU @ 2.60GHz (18-core), dual NVIDIA GeForce GTX 1080 Ti, 
and 32GB DDR4-RAM.	

As explained in Section~\ref{Architecture_Diversity}, we 
use \emph{DeepSpeech v0.1.0}~\cite{OnlineRef:DeepSpeech-Github-Repo} (called \texttt{DS0}) as the target model. 
 \emph{DeepSpeech v0.1.1}~\cite{OnlineRef:DeepSpeech-Github-Repo} (called \texttt{DS1}), \emph{Google Cloud Speech}~\cite{OnlineRef:Google-Cloud-Speech} (called \texttt{GCS}), and \emph{Amazon Transcribe}~\cite{OnlineRef:AmazonTranscribe} (called \texttt{AT}) are the three auxiliary ASRs. 
%The first two auxiliary ASRs are on-line services, while the last one runs locally with an officially pre-trained model. 
We use \texttt{X}$+$\{\texttt{Y$_1$}, ..., \texttt{Y$_n$}\} to
denote a system using \texttt{X} as the target model and \texttt{Y$_1$}, ..., \texttt{Y$_n$} as the auxiliary models.
%When only one auxiliary model is used, $n = 1$.
For example, \texttt{DS0}$+$\{\texttt{DS1}\} means
a system using \texttt{DS0} as the target model and \texttt{DS1} as the single
auxiliary model.

%The main reason
%we select \texttt{DS0} is that 
%its model architecture and parameters 
%are publicly available---making it
%possible to generate white-box AEs~\cite{DBLP:journals/corr/abs-1805-07820}.  
%But our system should also work if any other ASR model is selected as the target model. 

\begin{table}%[!htbp]
\normalsize
	\caption{Datasets used in our evaluation.}
	%\af
	\label{table:experiment-datasets}
	\centering
	\begin{tabular}{c|c|c} \hline
		\multicolumn{2}{c|}{Dataset Name}    & \# of Samples   \\ 
		\hhline{==|=}
		\multicolumn{2}{c|}{\emph{Benign}}  & 2400         \\  \hline                                         
		\multirow{2}*{\emph{AE}}  & White-box AEs & 1800                  \\ 
		  & Black-box AEs & 600                  \\ \hline
		   %& Part 3 (\emph{stress-test})  & 50                  \\ \hline
	\end{tabular}
	%\vspace{-6pt}
\end{table}

\subsection{Dataset Preparation} \label{subsec:dataset}

We consider two audio AE generation techniques:
white-box based~\cite{DBLP:conf/sp/Carlini018} and 
black-box based~\cite{DBLP:journals/corr/abs-1805-07820} methods,
and build two datasets: 
a \emph{Benign} dataset and an \emph{AE} 
dataset, each of which contains 2400 audio samples, as shown in 
Table~\ref{table:experiment-datasets}.
The audio samples of the \emph{Benign} dataset are randomly selected from the \emph{dev\_clean} dataset of LibriSpeech \cite{OnlineRef:LibriSpeech}.

The \emph{AE} dataset consists of the following two parts. (1)
1800 white-box AEs, including 990 AEs provided by \cite{DBLP:conf/sp/Carlini018} and 810 created by us. \todo{We created the AEs following the style of \cite{DBLP:conf/sp/Carlini018} by using randomly selected samples from
LibriSpeech as the host audios to embed the English sentences provided by \cite{DBLP:conf/sp/Carlini018}.} (2) 600 black-box AEs constructed by applying the black-box approach \cite{DBLP:journals/corr/abs-1805-07820};
each AE embeds only two words. \todo{The two-word limit is due to the current capacity of \cite{DBLP:journals/corr/abs-1805-07820}.}
\todo{Each white-box AE takes 18 minutes on average to generate, while each black-box
AE takes 90 minutes. The datasets and the transcription details are made
publicly available.\footnote{\url{https://goo.gl/CJmrQh}}}
We have verified that
all AEs can successfully fool the target model \texttt{DS0}.

%as the similarity score of the \emph{stress-test} AE and its host speeches should be high and could bypass our detection more easily compared to the AEs of other two part.

%, so we consider such cases as extreme cases and would likt to evaluate whether our proposed system work well on the extreme cases or not. 

%For each dataset, 80\% of its audio samples are used for training, and
%the remaining 20\% for testing.

\begin{figure*}%[!htbp]
	\centering
	\subfloat[\texttt{DS0}$+$\{\texttt{DS1}\}]{\includegraphics[width=0.33\textwidth]{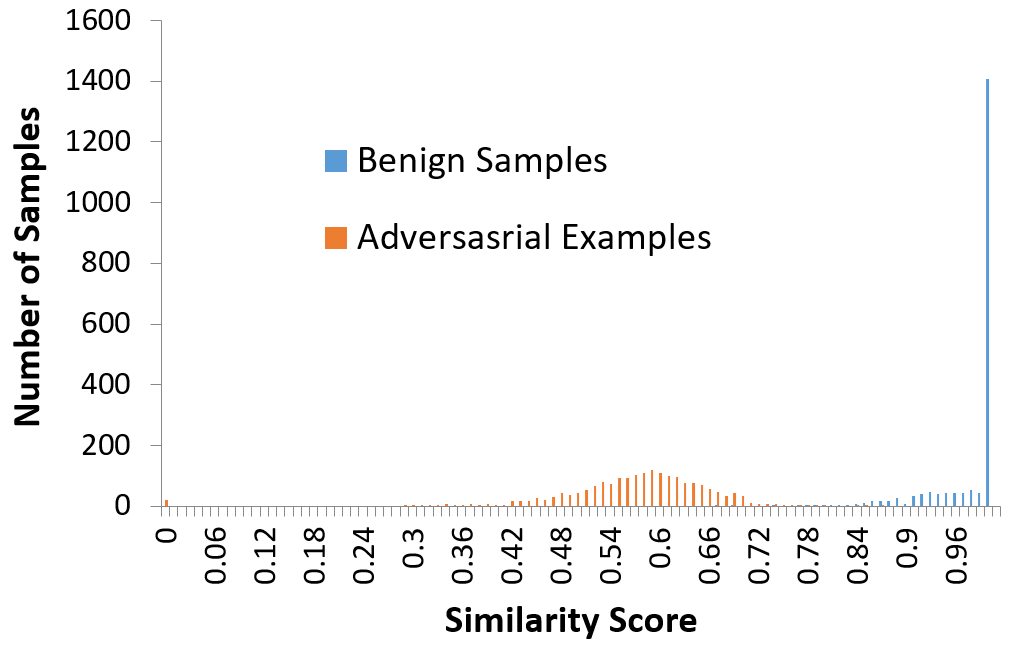}}~
	\subfloat[\texttt{DS0}$+$\{\texttt{GCS}\}]{\includegraphics[width=0.33\textwidth]{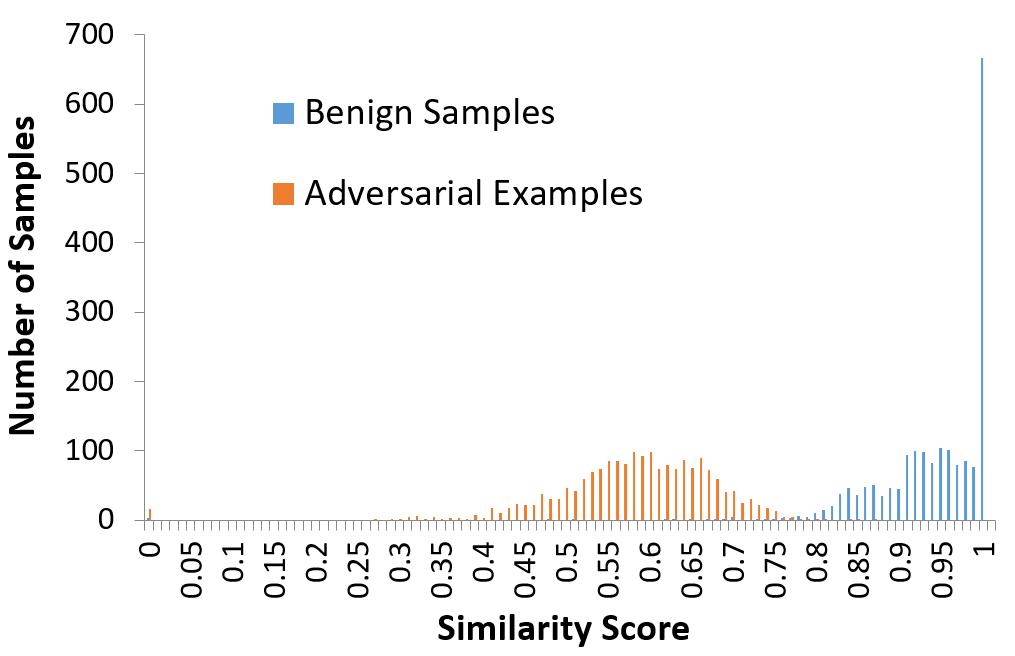}}~
	\subfloat[\texttt{DS0}$+$\{\texttt{AT}\}]{\includegraphics[width=0.33\textwidth]{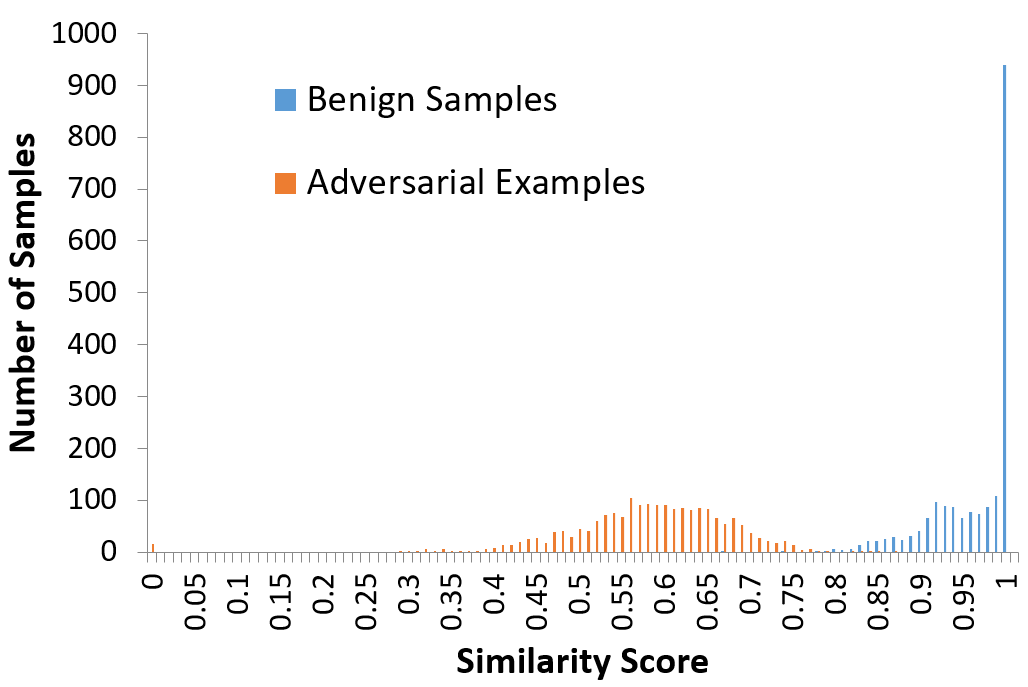}}~
	%\vspace{3pt}
	\caption{Similarity score histograms of the three \emph{single}-auxiliary-model systems.}
	\label{fig:distribution_of_similarity_scores}
\end{figure*}

%To study the feasibility of our system, we seek to understand whether it could 
%provide high similarity scores to the benign samples and low similarity scores 
%to AEs, such that benign samples and AEs could be separated well.
%(our intuition is that the similarity score of an AE should be low).

\subsection{Feasibility Analysis} \label{sec:feaibility}

To validate the feasibility of the MVP-inspired idea, we 
feed all the benign samples and AEs into each of the four ASRs,
and calculate the similarity scores. Given an audio (benign sample or AE),
the similarity score indicates the similarity of the transcription of \texttt{DS0} and
that of an auxiliary ASR.

The histograms in Figure~\ref{fig:distribution_of_similarity_scores} confirm
the feasibility of the idea: in each of the three cases, 
the similarity scores for AEs and those for benign samples 
form two almost disjoint clusters. It means that benign samples
lead to relatively high similarity scores, while AEs low scores.

%To this end, for each audio sample in the training dataset, we use 
%the \emph{phonetic encoding} technique and \emph{Jarro-Winkler} distance method 
%to calculate 
%the similarity score between the transcriptions generated by \texttt{DS0} and that by 
%one of the three auxiliary ASRs (i.e., \texttt{DS1}, \texttt{GCS} 
%and \texttt{AT}). 
%As seen in Figure~\ref{fig:distribution_of_similarity_scores}, the similarity scores for AEs and those for benign samples form two clusters.  

\begin{table*}[t]
	\centering
	\normalsize
	\caption{Accuracies when different similarity calculation methods are used.}
	%\af
	\label{Table:Performance Result-Comparison of 6 Similarity Metrics}
	%\resizebox{\linewidth}{!}{%
	\scalebox{0.96}{
	\begin{tabular}{|c|c|c|c|c|c|}
		\hline  
		\multirow{2}*{Similarity Method}  & \multirow{2}*{Performance} &  \multicolumn{4}{c|}{System} \\
		\cline{3-6}
		& &	\texttt{DS0}$+$\{\texttt{DS1}, \texttt{GCS}\} & \texttt{DS0}$+$\{\texttt{DS1}, \texttt{AT}\} & \texttt{DS0}$+$\{\texttt{GCS}, \texttt{AT}\} & \texttt{DS0}$+$\{\texttt{DS1}, \texttt{GCS}, \texttt{AT}\}            \\ \hline
		\multirow{3}{*}{\emph{Cosine}} 		& Accuracy  & 958/960 (99.79\%) & 957/960 (99.69\%) & 952/960 (99.17\%) & 957/960 (99.69\%) \\ \cline{2-6}
		                                & FPR       & 2/480 (0.42\%)    & 2/480 (0.42\%)    & 7/480 (1.46\%)    & 2/480 (0.42\%)    \\ \cline{2-6}
		                                & FNR       & 0/480 (0.00\%)    & 1/480 (0.21\%)    & 1/480 (0.21\%)    & 1/480 (0.21\%)    \\ \hline		
		\multirow{3}{*}{\emph{Jaccard}} 		& Accuracy  & 957/960 (99.69\%) & 958/960 (99.79\%) & 921/960 (95.94\%) & 956/960 (99.58\%) \\ \cline{2-6}
		                                & FPR       & 3/480 (0.63\%)    & 2/480 (0.42\%)    & 38/480 (7.92\%)    & 4/480 (0.83\%)    \\ \cline{2-6}
		                                & FNR       & 0/480 (0.00\%)    & 0/480 (0.00\%)    & 1/480 (0.21\%)    & 0/480 (0.00\%)    \\ \hline
		\multirow{3}{*}{\emph{JaroWinkler}}	 & Accuracy  & 958/960 (99.79\%) & 958/960 (99.79\%) & 956/960 (99.58\%) & 958/960 (99.79\%) \\ \cline{2-6}
		                                & FPR       & 1/480 (0.21\%)    & 1/480 (0.21\%)    & 2/480 (0.42\%)    & 1/480 (0.21\%)    \\ \cline{2-6}
		                                & FNR       & 1/480 (0.21\%)    & 1/480 (0.21\%)    & 2/480 (0.42\%)    & 1/480 (0.21\%)    \\ \hline
		\multirow{3}{*}{\emph{PE\_Cosine}} 	& Accuracy  & 958/960 (99.79\%) & 958/960 (99.79\%) & 957/960 (99.69\%) & 958/960 (99.79\%) \\ \cline{2-6}
		                                & FPR       & 2/480 (0.42\%)    & 2/480 (0.42\%)   & 2/480 (0.42\%)    & 2/480 (0.42\%)    \\ \cline{2-6}
	                                	& FNR       & 0/480 (0.00\%)    & 0/480 (0.00\%)    & 1/480 (0.21\%)    & 0/480 (0.00\%)    \\ \hline
		\multirow{3}{*}{\emph{PE\_Jaccard}} 	& Accuracy  & 958/960 (99.79\%) & 958/960 (99.79\%) & 443/960 (99.69\%) & 958/960 (99.79\%) \\ \cline{2-6}
	                                	& FPR       & 2/480 (0.42\%)    & 2/480 (0.42\%)    & 3/480 (0.63\%)    & 2/480 (0.42\%)    \\ \cline{2-6}
		                                & FNR       & 0/480 (0.00\%)    & 0/480 (0.00\%)    & 0/480 (0.00\%)    & 0/480 (0.00\%)    \\ \hline
		\multirow{3}{*}{\emph{PE\_JaroWinkler}}	 & Accuracy  & 958/960 (\textbf{99.79\%}) & 959/960 (\textbf{99.90\%}) & 958/960 (\textbf{99.79\%}) & 959/960 (\textbf{99.90\%}) \\ \cline{2-6}
		                                & FPR       & 1/480 (0.21\%)    & 0/480 (0.00\%)    & 0/480 (0.00\%)    & 1/480 (0.21\%)    \\ \cline{2-6}
		                                & FNR       & 1/480 (0.21\%)    & 1/480 (0.21\%)    & 2/480 (0.42\%)    & 0/480 (0.00\%)    \\ \hline																		
	\end{tabular}
	}
	%\vspace{-6pt}
\end{table*}

\subsection{Choosing Similarity Calculation Methods} \label{sec:metrics}
\label{subsection:discussion_similarity_metrics}

\todo{Many methods can be used to measure the similarity between two strings.
Some well-known ones include
\emph{Jaccard index}~\cite{OnlineRef:Jaccard-Similarity}, 
\emph{Cosine similarity}, %~\cite{OnlineRef:Cosine-Similarity}, 
and \emph{Edit distance} (e.g., \emph{JaroWinkler}~\cite{OnlineRef:Jaro-Winkler-Similarity}).
In addition, we propose to convert the transcription into
its phonetic encoding (PE)~\cite{OnlineRef:PhoneticEncoding}. 
Our hypothesis is that this may help reduce the false positives:
given a benign sample, different ASRs may output different words of
similar pronunciations, which can lead to  higher similarity scores
after phonetic encoding.}

\todo{To choose the similarity measurement method and validate the
hypothesis above, we consider six different combinations of
similarity calculation methods, as shown in
Table~\ref{Table:Performance Result-Comparison of 6 Similarity Metrics}.
For example, \emph{PE\_JaroWinkler} means phonetic encoding and
JaroWinkler are used as the similarity calculation method. 
In each case, we train a SVM based classifier 
using 80\% of audios in the two datasets (Benign and AE), and then
test the classifier using the rest 20\% (that is, 4800 * 0.2 = 960 audios). 
Table~\ref{Table:Performance Result-Comparison of 6 Similarity Metrics}
shows that the combination \emph{PE\_JaroWinkler}
achieves the highest accuracies in all the four example systems.
For example, when \texttt{DS1}, \texttt{GCS}, \texttt{AT}\ are
used as the auxiliary ASRs, it achieves the accuracy 99.90\%(,
although other methods also lead to high accuracies at least 99.58\%).
Thus, we choose \emph{PE\_JaroWinkler} 
as the similarity calculation method. 
}

\begin{table*}[!htbp]
	\centering
	\normalsize

\vspace{6pt}	\caption{Testing results of \emph{single}-auxiliary-model systems.
The mean and standard deviation across the 5-fold cross validation are reported, 
denoted as \texttt{mean/STD}. FPR: false positive rate; FNR: false negative rate.}
	%\af
	\label{Table:Results-of-One-Auxiliary-Model Systems Upon Testing Dataset}
	\scalebox{0.96}{
	\begin{tabular}{|c|c|c|c|c|}
		\hline
		\multirow{2}*{Classifier}  & \multirow{2}*{Performance} &  \multicolumn{3}{c|}{System} \\
		\cline{3-5}
		& & \texttt{DS0}$+$\{\texttt{DS1}\} & \texttt{DS0}$+$\{\texttt{GCS}\} & \texttt{DS0}$+$\{\texttt{AT}\}    \\ 
		\hhline{|=|=|=|=|=|}

		\multirow{3}{*}{\texttt{SVM}} & Accuracy    & 99.56\% / 0.18\%   & 98.92\% / 0.22\%   & 99.71\% / 0.14\%   \\ \cline{2-5}
		                                & FPR       & 0.38\% / 0.16\%    & 1.71\% / 0.40\%    & 0.25\% / 0.24\%    \\ \cline{2-5}
	                                	& FNR       & 0.50\% / 0.41\%    & 0.46\% / 0.24\%    & 0.34\% / 0.21\%    \\ \hline
		\multirow{3}{*}{\texttt{KNN}} & Accuracy    & 99.36\% / 0.12\%   & 98.35\% / 0.12\%   & 99.65\% / 0.16\%   \\ \cline{2-5}
		                                & FPR       & 0.67\% / 0.27\%    & 2.04\% / 0.16\%    & 0.25\% / 0.24\%    \\ \cline{2-5}
		                                & FNR       & 0.63\% / 0.48\%    & 1.25\% / 0.19\%    & 0.46\% / 0.24\%    \\ \hline
		\multirow{3}{*}{\texttt{Random Forest}} & Accuracy  & 99.31\% / 0.19\% & 98.04\% / 0.15\% & 99.54\% / 0.21\%  \\ \cline{2-5}
		                                & FPR       & 0.63\% / 0.23\%    & 2.21\% / 0.39\%    & 0.46\% / 0.24\%    \\ \cline{2-5}
		                                & FNR       & 0.75\% / 0.60\%    & 1.71\% / 0.15\%    & 0.46\% / 0.33\%    \\ \hline							
	\end{tabular}
	}
\end{table*}

\subsection{Effectiveness of Single-Auxiliary-Model Systems} \label{sec:single-model}

\todo{A detection system that uses a single auxiliary ASR 
has the advantage of a cheap deployment. We build three such systems:
\texttt{DS0}$+$\{\texttt{DS1}\}, \texttt{DS0}$+$\{\texttt{GCS}\}, and 
\texttt{DS0}$+$\{\texttt{AT}\}. 
We use k-fold cross validation (k = 5) to evaluate the three systems:
the datasets are divided into 5 equal subsets and, in each of the five runs,
one subset is used for testing and the rest four for training. 
As shown in Table~\ref{Table:Results-of-One-Auxiliary-Model Systems Upon Testing Dataset}, the mean and standard deviation of the results
across the 5 runs are reported.}

\todo{We use three different binary classifiers, including \texttt{SVM}, \texttt{KNN} and \texttt{Random Forest}, and configure each classifier as follows: (1) \texttt{SVM} uses a 3-degree polynomial kernel; (2) \texttt{KNN} uses 10 neighbors to vote; and (3) \texttt{Random Forest} uses a seed of 200 as the starting random state.}

\todo{Based on the results, we conclude that (1) all the single-auxiliary-model systems
achieve high accuracies (over 98\%) and low FPRs/FNRs; and (2) 
SVM performs slightly better than the other two classifier methods.}

\todo{But it is worth mentioning that if the auxiliary ASR (like \emph{Kaldi}) 
is not accurate in recognizing benign audios, the AE detection accuracy is
bad (e.g., \textless 80\% with \emph{Kaldi})}.

\begin{table*}[!htbp]
	\centering
	\normalsize

\vspace{6pt}	\caption{5-fold cross validation testing results (reported as \texttt{mean/STD}) of four \emph{multi}-auxiliary-model systems.}
	%\af
	\label{Table:Results of Multiple-Auxiliary-Model Systems Upon Testing Dataset}
	\scalebox{0.96}{
	\begin{tabular}{|c|c|c|c|c|c|}
		\hline
		\multirow{2}*{Classifier}  & \multirow{2}*{Performance} &  \multicolumn{4}{c|}{System} \\
		\cline{3-6}
		& & \texttt{DS0}$+$\{\texttt{DS1}, \texttt{GCS}\} & \texttt{DS0}$+$\{\texttt{DS1}, \texttt{AT}\} & \texttt{DS0}$+$\{\texttt{GCS}, \texttt{AT}\} & \texttt{DS0}$+$\{\texttt{DS1}, \texttt{GCS}, \texttt{AT}\}     \\ 
		\hhline{|=|=|=|=|=|=|}
%Classifier & Performance & DS1+GCS+AT        & DS1+GCS           & DS1+AT            & GCS+AT            \\ \hline
		\multirow{3}{*}{\texttt{SVM}} & Accuracy    & 99.75\% / 0.05\%   & 99.86\% / 0.08\%   & 99.82\% / 0.10\%   & 99.88\% / 0.10\%  \\ \cline{2-6}
		                                & FPR       & 0.29\% / 0.21\%    & 0.08\% / 0.10\%    & 0.08\% / 0.10\%    & 0.04\% / 0.08\%   \\ \cline{2-6}
	                                	& FNR       & 0.21\% / 0.23\%    & 0.21\% / 0.23\%    & 0.29\% / 0.21\%    & 0.21\% / 0.23\%    \\ \hline
		\multirow{3}{*}{\texttt{KNN}} & Accuracy    & 99.77\% / 0.04\%   & 99.81\% / 0.08\%   & 99.75\% / 0.17\%   & 99.86\% / 0.08\% \\ \cline{2-6}
		                                & FPR       & 0.25\% / 0.16\%    & 0.13\% / 0.10\%    & 0.21\% / 0.23\%    & 0.08\% / 0.10\%    \\ \cline{2-6}
		                                & FNR       & 0.21\% / 0.23\%    & 0.25\% / 0.21\%    & 0.29\% / 0.21\%    & 0.21\% / 0.23\%   \\ \hline
		\multirow{3}{*}{\texttt{Random Forest}} & Accuracy  & 99.73\% \ 0.08\%  & 99.81\% / 0.12\% & 99.77\% / 0.08\% & 99.84\% / 0.08\% \\ \cline{2-6}
		                                & FPR       & 0.25\% / 0.16\%    & 0.13\% / 0.17\%    & 0.17\% / 0.08\%    & 0.08\% / 0.10\%    \\ \cline{2-6}
		                                & FNR       & 0.29\% / 0.28\%    & 0.25\% / 0.21\%    & 0.29\% / 0.21\%    & 0.25\% / 0.21\%    \\ \hline							
	\end{tabular}
	}
\aaf
\end{table*}

\begin{table}[h]
\centering
\caption{Impact of the number of ASRs on FPR and FNR.}
\label{Table:impackt_of_the_number_of_ASRs_on_FPR_and_FNR}
\begin{tabular}{|c|c|c|c|}
\hline
\# of Aux.\ ASRs & System               & FPR   & FNR            \\ \hline
\multirow{3}*{1}    & DS0+\{DS1\}          & 0.38\%  & 0.50\% \\
                    & DS0+\{GCS\}          & 1.71\%  & 0.46\% \\
                    & DS0+\{AT\}           & 0.25\%  & 0.34\% \\ \hline
\multirow{3}*{2}    & DS0+\{DS1, GCS\}     & 0.29\%  & 0.21\% \\ 
                    & DS0+\{DS1, AT\}      & 0.08\%  & 0.21\% \\ 
                    & DS0+\{GCS, AT\}      & 0.08\%  & 0.29\% \\ \hline
3                   & DS0+\{DS1, GCS, AT\} & 0.04\%  & 0.21\% \\ \hline
\end{tabular}
\end{table}

\subsection{Effectiveness of Multi-Auxiliary-Model Systems} \label{sec:multiple-model}
A \emph{multi-auxiliary-model} system uses 
\emph{more than one} auxiliary model. We 
build four multiple-auxiliary-model systems, denoted as:
\texttt{DS0}$+$\{\texttt{DS1}, \texttt{GCS}\},
\texttt{DS0}$+$\{\texttt{DS1}, \texttt{AT}\},
\texttt{DS0}$+$\{\texttt{GCS}, \texttt{AT}\},
and \texttt{DS0}$+$\{\texttt{DS1}, \texttt{GCS}, \texttt{AT}\}.
%each of which uses \texttt{DS0} as the target model, and the ones 
%within a bracket as the auxiliary models. For example, 
%\texttt{DS0}$+$\{\texttt{DS1}, \texttt{GCS}\} has two auxiliary models, \texttt{DS1} and \texttt{GCS}.

For a multi-auxiliary-model system with $n$ auxiliary models, 
given an input audio, $n$ similarity 
scores are calculated and form a feature vector. 
The feature vector is then fed into the binary 
classifier to predict whether the audio is an AE. % (assume the binary classifier is already trained). 

\todo{Table~\ref{Table:Results of Multiple-Auxiliary-Model Systems Upon Testing Dataset} shows the testing results using 5-fold cross validation. All the accuracy results are
higher than 99.70\%, and FPR and FNR are lower than 0.30\%, regardless of 
the auxiliary models and binary classifiers used. 
The three-auxiliary-model system performs the best and reaches the accuracy
99.88\%. We also observe that a system with more auxiliary models achieves
better accuracy, probably because extra models provide more features in 
the feature vector.}

\noindent \textbf{Does FPR increase due to more auxiliary ASRs?}
\todo{To answer this intriguing question, we extract the FPR and
FNR results when SVM is used as the binary classifier from Table \ref{Table:Results-of-One-Auxiliary-Model Systems Upon Testing Dataset} and Table \ref{Table:Results of Multiple-Auxiliary-Model Systems Upon Testing Dataset} and obtain Table \ref{Table:impackt_of_the_number_of_ASRs_on_FPR_and_FNR},
which shows that both FPRs and FNRs tend to slightly decline when more
auxiliary ASRs are used. This conclusion holds when either 
KNN or RandomForest is used as the binary classifier.}

%For each of MAM systems, we train a binary classifier on all the 1800 feature vectors. Then each trained model is tested over the 450 test samples. In order to see the possible performance differences between scenarios of using different binary classifiers, SVM, KNN and Random Forest are compared in our experiment. Training codes of three classifiers are obtained from sklearn \cite{OnlineRef:sklearn-Python-Package}. And detailed configuration for each classifier is as follow: 1) SVM uses a 3-degree polynomial kernel, 2) KNN uses 10 neighbors to vote, 3) Random Forest uses a seed of 200 as starting random state. Results are listed in Table \ref{Table:Results of Multiple-Auxiliary-Model Systems Upon Testing Dataset}. All experiment cases reach a high accuracy of more than 99\%, and low FPR and FNR, each of which is lower than 0.5\%. The result also shows a trend that the Three-Auxiliary-Models system perform better than any of these two-auxiliary-models systems. We credit this improvement to the extra information brought by a 3rd auxiliary model.

\begin{table}[t]
	\centering
	\vspace{3pt}
	\normalsize
	\caption{The detection results of unseen-attack AEs for three \emph{single}-auxiliary-models.}
	%\af
	\label{Table:unknown-AEs-attack-on-1-aux-model-systems}
	\scalebox{0.87}{
	\hspace{-2mm}
	\begin{tabular}{|c|c|c|c|c|c|}
		\hline
		System & Threshold & FPR    & FNs & FNR    & Defense rate \\ \hline
		\texttt{DS0}$+$\{\texttt{DS1}\}        & 0.88      & 4.13\% & 0  & 0.00\% & 100\%         \\ \hline
		\texttt{DS0}$+$\{\texttt{GCS}\}        & 0.82      & 4.75\% & 4  & 0.17\% & 99.83\%         \\ \hline
		\texttt{DS0}$+$\{\texttt{AT}\}         & 0.85      & 3.92\% & 2  & 0.08\% & 99.92\%         \\ \hline
	\end{tabular}
	}
	\vspace{3pt}
\end{table}

\begin{figure*}[!htbp]
	\centering
	\subfloat[\texttt{DS0}$+$\{\texttt{DS1}\}]{\includegraphics[width=0.33\textwidth]{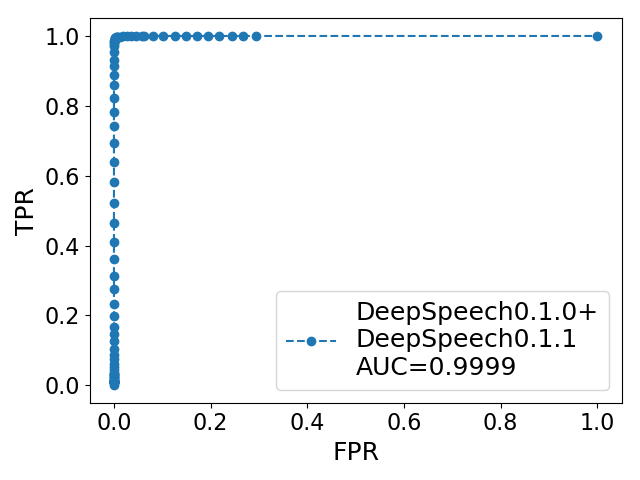}}~
	\subfloat[\texttt{DS0}$+$\{\texttt{GCS}\}]{\includegraphics[width=0.33\textwidth]{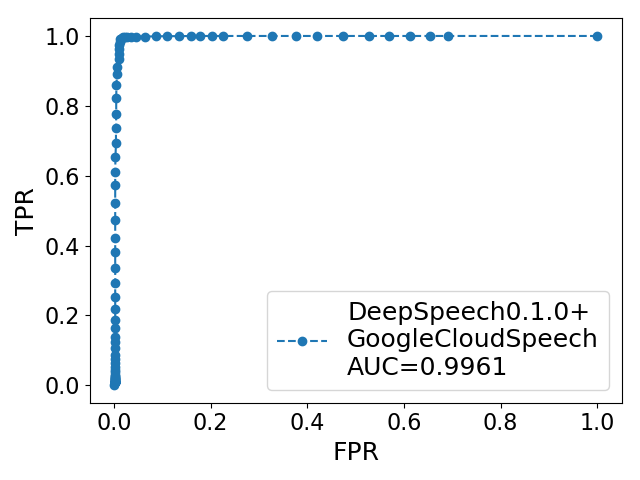}}~
	\subfloat[\texttt{DS0}$+$\{\texttt{AT}\}]{\includegraphics[width=0.33\textwidth]{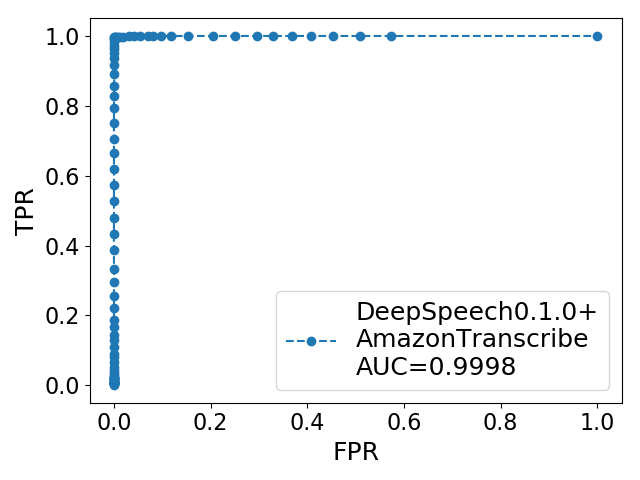}}~
	\caption{The ROC curves of the three \emph{single}-auxiliary-model systems.}
	\label{fig:3_RoC_curves}
	%\aaf\af
\end{figure*}

\subsection{Robustness against Unseen Attack Methods} \label{sec:robustness}

This experiment aims to examine whether a system trained on AEs generated by a particular attack method is able to detect AEs generated by other kinds of attack methods---such an AE is called an \emph{unseen-attack AE}.  
%evaluates the robustness of the systems against AEs generated by \emph{unseen} attack methods---AEs that have not been seen during training. 
We use the \emph{defense rate}, defined as the ratio of the number of successfully 
detected AEs among the total number of AEs, to measure
the robustness. 
%The higher defense rate, the better robustness of the system. 

\vspace{3pt}
\noindent
\textbf{Single-auxiliary-model systems.}
We first examine all the three single-auxiliary-model systems.
We train each system using only the 2400 benign samples, and test it  %single-auxiliary-model 
using all the 2400 AEs (see Table~\ref{table:experiment-datasets}),
which all can can be considered as unseen-attack AEs.

We simply use a similarity score threshold $T$ for detecting AEs:
an audio that has a score lower than $T$ is detected as an AE.
First, the threshold is determined by having the FPR less than 5\%.
The detection results are presented in 
Table~\ref{Table:unknown-AEs-attack-on-1-aux-model-systems}.
%For each system, the threshold is determined by having the FPR less than 5\%. 
Each of the three detection systems achieves excellent defense rates 
$\ge$ 99.83\%. 
Second, we vary $T$ and obtain the Receiver Operating Characteristic (ROC) 
curve, as shown in Figure~\ref{fig:3_RoC_curves}.
The AUC in each case is close to 1, implying high detection accuracies,
which is consistent with the results of the machine learning based approaches (see Section~\ref{sec:single-model}).
%We further try different upper bound values for FPR, including
%4\%, 3\% and 2\% and still obtain high defense rates. E.g., the defense rates of \texttt{DS0}$+$\{\texttt{DS1}\}, \texttt{DS0}$+$\{\texttt{GCS}\} and \texttt{DS0}$+$\{\texttt{AT}\} slightly drop to 99.92\%, 99.54\% and 99.83\%, respectively, 
%when 2\% is used as as the maximum FPR.

%\todo{drop 2\%, but the lowest defense rate is still 98.49\% ??}

\begin{table}[t]
	\centering
	\caption{The detection results of unseen-attack AEs for four multiple-auxiliary-models.}
    %\af
	\label{Table:unknown-AEs-attack-on-multiple-aux-model-systems}
	\resizebox{\linewidth}{!}{%
	\begin{tabular}{|c|c|c|}
		\hline
		\multirow{2}{*}{System} & \multicolumn{2}{c|}{Defense rate}  \\
		\cline{2-3}
		& Black-box AEs & White-box AEs \\ \hline
\texttt{DS0}$+$\{\texttt{DS1}, \texttt{GCS}\}   & 99.33\%                   & 100\%                     \\ \hline
\texttt{DS0}$+$\{\texttt{DS1}, \texttt{AT}\}   & 99.17\%                   & 100\%                     \\ \hline
\texttt{DS0}$+$\{\texttt{GCS}, \texttt{AT}\}   & 99.33\%                   & 99.89\%                     \\ \hline
\texttt{DS0}$+$\{\texttt{DS1}, \texttt{GCS}, \texttt{AT}\}	 & 99.33\%                   & 100\%                     \\ \hline
	\end{tabular}
	}
\end{table}

\vspace{3pt}
\noindent
\textbf{Multiple-auxiliary-model systems.}
We next examine the four multiple-auxiliary-model systems. 
As aforementioned, two different methods 
are used to generate AEs:
the white-box approach and black-box approach. 
We have totally 1800 white-box AEs and 600 black-box AEs.
%(all \emph{stree-test} AEs are white-box AEs). 
We conduct two different experiments to evaluate each of the four  
multiple-auxiliary-model systems.

We first use all the 1800 white-box AEs and 1800 benign samples to train 
each system, and use all the black-box AEs to test each trained system. 
The results are showed in the second column in Table~\ref{Table:unknown-AEs-attack-on-multiple-aux-model-systems}.
We can see that all the systems perform very well, and the lowest defense 
rate is 99.17\% for the \texttt{DS0}$+$\{\texttt{AT}\} system.

We next use all the 600 black-box AEs and 600 benign samples to train each system, and use all the white-box AEs to test each trained system. 
The results are showed in the third column in Table~\ref{Table:unknown-AEs-attack-on-multiple-aux-model-systems}.
There are three systems achieve the defense rate of 100\%---that is,
all the white-box AEs can be successfully detected. For the system  \texttt{DS0}$+$\{\texttt{GCS}, \texttt{AT}\}, it achieve a high defense 
rate of 99.89\%.

Therefore, we can conclude that our detection systems are very robust against
unseen-attack AEs.

\subsection{Detecting Hypothetical Multiple-ASR-Effective AEs}\label{subsec:hypothetical_multiple-ASRs-effective_AE}

In this experiment, we seek to understand whether our proposed system can work well when facing AEs that can fool more than one ASRs, which we call \emph{multiple-ASR-effective (MAE) AEs}. 
The problem here is the lack of methods for generating such MAE AEs. We thus propose to create \emph{hypothetical} MAE AEs. Note that we do not really create a real MAE AE in the form of an audio, instead we synthesize a feature vector to represent it.
In the following presentation, we refer to a \emph{hypothetical MAE AE} as a \emph{MAE AE}. 

%in the feature vector, each element is a similarity score representing the similarity between the MAE AE's transcription generated by the target ASR and that by an auxiliary ASR. 

\vspace{3pt}
\noindent
\textbf{Creating MAE AEs.}
We use a multiple-auxiliary-model system, \texttt{DS0}$+$\{\texttt{DS1}, \texttt{GCS}, \texttt{AT}\}, in this experiment. Thus, for a given input audio, 
its feature vector contains three similarity scores.  
\todo{We have the following \emph{critical observation}.
If an MAE AE successfully fools the target model and an auxiliary model, 
both models should convert the audio into transcriptions the same as or highly
similar to the 
command desired by the attacker (since it is a targeted attack). 
This AE works just like a benign sample (whose transcription is the command)
for the perspective of the two models. Thus, the similarity
score of the AE corresponding to the auxiliary model should be 
as high as that of a benign sample. We thus construct the feature
vectors for MAE AEs as follows.} 

First, from the previous experiments, we collect two pools of similarity scores: one contains the similarity scores for the 2400 benign samples, denoted as $\lambda_{Be}$; and the other contains the similarity scores for the 2400 AEs, denoted as $\lambda_{Ak}$. 
%Note that all the AEs \emph{by default} can attack the target model \texttt{DS0}. 
%, but cannot attack the three auxiliary models.

%Note that all the AEs can successfully fool \texttt{DS0}.  

Second, to create an MAE AE that can fool \texttt{DS0} and \texttt{DS1}, for example, we randomly select a similarity score from $\lambda_{Be}$ (representing that this AE can successfully attack \texttt{DS0} and \texttt{DS1}), and two similarity scores from $\lambda_{Ak}$ (representing that this AE cannot attack \texttt{GCS} and \texttt{AT}, resulting in a low similarity score for each ASR). The created MAE AE is denoted as 
$AE(\texttt{DS0},\texttt{DS1})$. 

Similarly, consider creating an MAE AE that can fool \texttt{DS0}, \texttt{DS1}, and \texttt{GCS}, as
another example. This AE is denoted as 
$AE(\texttt{DS0},\texttt{DS1},\texttt{GCS})$.
Two similarity scores are selected from $\lambda_{Be}$ (representing that this AE can successfully attack \texttt{DS0}, \texttt{DS1}, and \texttt{GCS}), and one similarity score from $\lambda_{Ak}$ (representing that this AE cannot attack \texttt{AT}).

\begin{table}
\vspace{3pt}
\centering
\normalsize
\caption{Six different types of \emph{hypothetical} MAE AEs.}
%\af
\label{Table:6_hypothetical_multiple-ASRs-effective_AEs}
\scalebox{0.96}{
\begin{tabular}{|c|c|c|}
\hline
Type  & MAE AE           & \# of MAE AEs \\ 
\hhline{|=|=|=|}
%1-ASR-effective                   & AE\{DS0\}           & DS0             \\ \hline
\emph{Type-1} & $AE(\texttt{DS0},\texttt{DS1})$    & 2,400        \\ \hline 
\emph{Type-2} & $AE(\texttt{DS0},\texttt{GCS})$    & 2,400        \\ \hline 
\emph{Type-3} & $AE(\texttt{DS0},\texttt{AT})$     & 2,400         \\ \hline
\emph{Type-4} & $AE(\texttt{DS0},\texttt{DS1},\texttt{GCS})$ & 2,400   \\ \hline 
\emph{Type-5} & $AE(\texttt{DS0},\texttt{DS1},\texttt{AT})$  & 2,400    \\ \hline 
\emph{Type-6} & $AE(\texttt{DS0},\texttt{GCS},\texttt{AT})$  & 2,400    \\ \hline
\end{tabular}
}
\end{table}

Through this, we create six different types of MAE AEs, as listed in Table \ref{Table:6_hypothetical_multiple-ASRs-effective_AEs}. Each type contains 2400 MAE AEs.

\vspace{3pt}
\noindent
\textbf{Accuracy.}
For each type of MAE AEs, we construct six datasets:
each dataset contains 2400 benign samples and 2400 the corresponding MAE AEs.
For each dataset, 80\% of its benign samples and MAE AEs are used for training, and the remaning 20\% for testing. 
%The reason we include the original AEs is that we want to examine if the trained model can still detect original AEs. 
We use SVM as the binary classifier, and \emph{PE\_JaroWinkler} to measure the similarity. 
%Note that for an MAE AE we directly use its feature vector
%consisting of a list of similarity scores for training and testing. 

 \begin{table}
\centering
\normalsize
%\vspace{3pt}
\caption{Testing results of the system with respect to different types of MAE AEs.}
\label{Table:testing_result_detection_system_built_on_hypothetical_AEs}
%\af
\scalebox{0.96}{
\begin{tabular}{|c|c|c|c|}
\hline
MAE AE type            & Accuracy & FPR    & FNR    \\ 
\hhline{|=|=|=|=|}
%Detector\_AE\{DS0\}           & 99.33\%  & 1.17\% & 0.17\% \\ \hline
\emph{Type-1}      & 98.12\%  & 3.75\% & 0.00\% \\ \hline
\emph{Type-2}      & 99.25\%  & 1.50\% & 0.00\% \\ \hline
\emph{Type-3}      & 99.12\%  & 1.75\% & 0.00\% \\ \hline
\emph{Type-4}  & 96.46\%  & 5.34\% & 1.75\% \\ \hline
\emph{Type-5}  & 97.02\%  & 3.46\% & 2.50\% \\ \hline
\emph{Type-6}  & 98.52\%  & 2.46\% & 0.50\% \\ \hline
\end{tabular}
}
%\aaf\aaf
\end{table}
 
Table~\ref{Table:testing_result_detection_system_built_on_hypothetical_AEs} shows the testing results. We can see that the systems trained on different types of MAE AEs have very high accuracies (higher than 97\%), and low FPRs and FNRs.

\begin{table*}
\centering
\vspace{3pt}
%\normalsize
\caption{Defense rates against unseen-attack MAE AEs.}
%\af
\label{Table:defense_rates_of_detection_systems_built_on_hypothetical_AEs}
\renewcommand{\arraystretch}{1.1}
\scalebox{0.94}{
\begin{tabular}{|c|c|c|c|c|c|c|c|}
\hline
\multirow{2}{*}{AEs included in } & \multicolumn{7}{c|}{AEs included in \emph{testing} dataset}                                                           \\ \cline{2-8} 
\multirow{2}{*}{\emph{training} dataset}       & \multirow{1}{*}{Original} & \emph{Type-1} & \emph{Type-2}   & \emph{Type-3}  & \emph{Type-4} & \emph{Type-5} & \emph{Type-6} \\ \cline{3-8} 

    & \multirow{1}{*}{AEs} & $AE(\texttt{DS0},\texttt{DS1})$  & $AE(\texttt{DS0},\texttt{GCS})$   & $AE(\texttt{DS0},\texttt{AT})$  & $AE(\texttt{DS0},\texttt{DS1},\texttt{GCS})$ & $AE(\texttt{DS0},\texttt{DS1},\texttt{AT})$ & $AE(\texttt{DS0},\texttt{GCS},\texttt{AT})$ 

\\ \hhline{|=|=|=|=|=|=|=|=|}
Original AEs               &   ---  & 99.83\%        & 99.83\%          & 99.83\%        & 36.79\%             & 30.33\%            & 65.17\%            \\ \hline
$AE(\texttt{DS0},\texttt{DS1})$          & \textbf{99.96\%} & ---    & 99.96\%          & 100\%          & 89.12\%             & 75.75\%            & 63.33\%            \\ \hline
$AE(\texttt{DS0},\texttt{GCS})$          & \textbf{99.83\%} & 99.88\%        & ---      & 99.71\%        & 68.13\%             & {\underline{16.04\%}}            & 82.58\%            \\ \hline
$AE(\texttt{DS0},\texttt{AT})$           & \textbf{99.83\%} & 99.67\%        & 99.71\%          & ---    & {\underline{20.75\%}}              & 58.50\%            & 86.21\%            \\ \hline
$AE(\texttt{DS0},\texttt{DS1},\texttt{GCS})$     & \textbf{100\%}   & \textbf{\emph{100\%}} & \textbf{\emph{100\%}}   & 100\%          & ---       & 76.17\%            & 74.38\%            \\ \hline
$AE(\texttt{DS0},\texttt{DS1},\texttt{AT})$      & \textbf{99.92\%} & \textbf{\emph{100\%}} & 99.67\%          & \textbf{\emph{100\%}} & 23.13\%             & ---      & 72.12\%            \\ \hline
$AE(\texttt{DS0},\texttt{GCS},\texttt{AT})$      & \textbf{99.92\%} & 89.62\%        & \textbf{\emph{99.96\%}} & \textbf{\emph{100\%}} & 10.08\%             & 16.04\%            & ---      \\ \hline
\end{tabular}
\aaf
\af
}

%\raggedright Note: AE\{DS0\} here represents actually generated AEs. Values along the diagonal indicate the TPR of the system against test set.
\end{table*}

\vspace{3pt}
\noindent
\textbf{Robustness to unseen-attack MAE AEs.}
We further investigate whether the system is able to detect unseen-attack MAE AEs.  
We also include the original AEs in this experiment. Specifically, we use one type of AEs 
and the benign samples to train the system, and use another type of AEs to test 
the trained system. The result is presented in Table \ref{Table:defense_rates_of_detection_systems_built_on_hypothetical_AEs}.

We can conclude that (1) all the systems trained on different types of AEs work very well on the original AEs---achieving more than 99\% defense rates (the second column in Table~\ref{Table:defense_rates_of_detection_systems_built_on_hypothetical_AEs}).

(2) If a system is trained on one type of MAE AEs that can fool a set of ASRs, $\Lambda=\{A_1,...,A_n\}$ where $A_i$ is an ASR, then it can defend against another type of MAE AEs that can fool a set of the ASRs, $\Lambda^\prime \subseteq \Lambda$, with almost 100\% defense rates (the \emph{italic} numbers in Table~\ref{Table:defense_rates_of_detection_systems_built_on_hypothetical_AEs}). For example, if the system is trained on \emph{Type-4} MAE AEs that can fool 
\texttt{DS0}, \texttt{DS1}, and \texttt{GCS}, then it can defend against \emph{Type-1} MAE AEs that can fool \texttt{DS0} and \texttt{DS1}.

%(3) For the system trained only on the original AEs, it works well on defending against MAE AEs that can fool only two ASRs (i.e., \emph{Type-1}, \emph{Type-2}, and \emph{Type-3} MAE AEs), but the performance is not so good when the system faces MAE AEs that can fool three ASRs, which is not surprising.

%(the third row in Table~\ref{Table:defense_rates_of_detection_systems_built_on_hypothetical_AEs}).

(3) We especially notice that some defense rates are quite low: for example, if the system is trained on \emph{Type-2} MAE AEs that can fool \texttt{DS0} and \texttt{GCS}, its defense rate against \emph{Type-5} MAE AEs that can fool \texttt{DS0}, \texttt{DS1} and \texttt{AT} is only 16.04\%. 
This is expected, as the training data has never seen \emph{Type-5} MAE AEs.
%This can be easily resolved by training a comprehensive classifier with 
%a dataset containing all the different types of MAE AEs.
%The reason is probably that although the system learns useful features for \emph{Type-2} MAE AEs, the features for \emph{Type-5} MAE AEs are quite different (no common feature is shared by the two types) and are not learned by the system.
%it may fail to extract useful features for MAE AEs during testing, while the training and testing MAE AEs do not share common features.
%But such a case is an extreme case. And in real deployment, we will not use the system trained on \emph{Type-2} MAE AEs to detect  \emph{Type-5} MAE AEs.
%Moreover, so far no systematic AE generation method exists that can generate \emph{real} transferable audio AEs effective across different ASR systems. %, not to mention three or more.
%that  system only learns useful features for detecting MAE AEs that can fool a set of ASRs, $\Lambda=\{A_1,...,A_n\}$, but has no 

%$\Lambda^\prime \cap \Lambda - \{\texttt{DS0}\} = \oldemptyset$

\begin{table}[t]
\centering
\caption{Defense rates of the comprehensive system against unseen-attack MAE AEs and the original AEs.}
\label{Table:defense_rates_of_the_comprehensive_system}
\begin{tabular}{|c|c|}
\hline
Unseen-attack AE & Defense rate   \\ \hline
Original AE      & 100\% \\ \hline
AE(DS0, DS1)     & 100\% \\ \hline
AE(DS0, GCS)     & 100\% \\ \hline
AE(DS0, AT)      & 100\% \\ \hline
\end{tabular}
\aaf
\end{table}

\vspace{3pt}
\noindent
\todo{
\textbf{Comprehensive system.}
To build a \emph{comprehensive system} that can defend against all the 6 types of MAE AEs, we build a system using Type-4, Type-5 and Type-6 MAE AEs.
They constitute a dataset of totally 7200 MAE AEs, and we also build another dataset
of (feature vectors corresponding to) 7200 benign samples. 
%For each of three types MAE AEs, 2400 synthesized feature vectors are used in the experiment. Correspondingly, 7200 feature vectors are synthesized to represent 7200 hypothetical benign samples and used in the experiment. For each of comprehensive AEs and hypothetical benign sample, their total 7200 feature vectors are randomly split into
We use 80\% of them for training and the rest for testing.
The detection result on the testing set (of the three types of MAE AEs and benign samples) shows 97.22\% detection accuracy, 3.47\% FPR and  2.08\% FNR. The system's 
defense rates over original AE, Type-1, Type-2 and Type-3 MAE AEs are impressive, as shown in Table \ref{Table:defense_rates_of_the_comprehensive_system}. All the four types of AEs are 100\% detected.
}

\todo{We thus conclude the MVP-inspired idea lets us proactively train systems that
can detect transferable AEs even before their existence. This is a big 
step ahead of attackers working on methods for generating transferable
audio AEs.}

\subsection{Overhead Measurement} \label{sec:overhead}
\todo{We evaluate the time overhead imposed by the detection.
Both Google Cloud Speech and Amazon Transcribe are remote cloud
based services, whose delays are dominated by networks and 
service dispatching. We are more interested in the overhead
imposed by the detection itself. Thus, we measure the time overhead
on \texttt{DS0}$+$\{\texttt{DS1}\}. We use SVM
as the binary classifier, and evaluate using all the
2400 AEs and 2400 benign samples in our datasets. 
The average recognition time of \texttt{DS0} 
is 8.8 seconds per audio. The time overhead in \texttt{DS0}$+$\{\texttt{DS1}\}
comprises three parts: recognition overhead (due to the parallel ASRs), 
similarity calculation overhead and  classification overhead,
which are 0.065s (0.74\%), 5.0e-06s and 4.2e-07s.
Thus, they incur negligible overheads. }

{\color{black}
\subsection{Detecting Non-targeted Attacks}\label{sec:untargeted}

Non-targeted AEs make the ASR system output 
incorrect transcriptions. Compared to targeted AEs, 
attacks based on non-targeted AEs are much weaker, as they cannot
fool the ASR system to generate attacker-desired results. 

%Most of existing works of targeted ASR adversarial examples generation also provide the option to generate untargeted AEs using the same technique as generating targeted Es~\cite{cisse2017houdini,khare2018adversarial,alzantot2018did,du2019sirenattack}. 
%There are very few pure untargeted attack methods found because they are difficult to pose a real threat~\cite{du2019sirenattack} and are much less effective than targeted methods~\cite{DBLP:conf/sp/Carlini018}. 

%\cite{alzantot2018did} can generate non-targeted AEs using their targeted AE generation methods,
%but it is not .
%Differently, they either choose of target text becomes random texts that have the largest word error rate (WER)~\cite{khare2018adversarial} or simply loose the evaluation metric to consider some unsuccessful targeted AEs that cannot lead to the desired transcript but can still mess up the output as successful untargeted AEs~\cite{DBLP:conf/sp/Carlini018,alzantot2018did}. \cite{du2019sirenattack,xu2018hasp} claimed they can create untargeted AEs that have transferability, but they neither open source their AE generation tool nor provide available AE samples. 
We find that non-targeted AEs actually can be generated by simply adding noises to 
benign audio samples. We randomly choose 118 benign audio samples from CommonVoice dataset~\cite{commonvoice} which are labeled with US accent and high-quality vote.
 The noise is added to the audio samples with SNR set to -6dB. After adding the noise, the output sample is still recognizable to humans but the ASR transcription  results have higher than 80\% word error rate (WER). 
 
We treat non-targeted AEs as unseen-attack AEs and use the same evaluation
method used in Section~\ref{sec:robustness}. Specifically, 
we choose the similarity score threshold by having the FPR as 5\%,
and use that threshold to determine whether an input is an AE.
The testing results show that the defense rate is higher than 90\%
no matter which auxiliary ASR is used. The defense rate is
lower than the case of targeted attack AEs, mainly
because of the  relatively small WER when generating non-targeted AEs.

%For the classifier is trained with either whitebox AEs and blackbox AEs. Still, we achieve a detection rate higher than 89\%.
}

\section{Related Work}\label{(Related-Work)}
\subsection{Audio Adversarial Example Generation}
%Attacks on ASRs can be classified according to the model transparency and the attacking directivity. 
An AE generation method can be categorized as whitebox/blackbox and targeted/non-targeted.
%As non-targeted audio AEs are trivial to generate and are , we omit discussing them. 
%Here we present a simple survey about existing targeted audio AE generation methods with both the whitebox and blackbox assumptions. 
Tavish \textit{et al.}~\cite{DBLP:conf/woot/VaidyaZSS15} and Carlini \textit{et al.}~\cite{DBLP:conf/uss/CarliniMVZSSWZ16} proposed methods for generating commands that are recognized 
by ASRs but are considered as noise by humans. However, the produced sound examples are noises and incomprehensible to humans, which greatly undermines the power of their attacks. 

To address this limitation, \cite{cisse2017houdini} proposed a semi-targeted AE generation method to embed text to host audios with similar content. 
Carlini proposed the first targeted audio AE generation system, but is is a whitebox method~\cite{DBLP:journals/corr/abs-1805-07820}.
Alzantot \textit{et al.} proposed a blackbox scheme to generate  adversarial audio examples targeting  a simple speech command classification model (not an ASR)~\cite{alzantot2018did}.
Taori et al.~\cite{DBLP:journals/corr/abs-1805-07820} combined the
genetic algorithm~\cite{alzantot2018did} and gradient estimation to 
generate blackbox AEs. Finally, Yuan \textit{et al.} aimed to craft audio AEs that remain effective when they are played over the air\cite{DBLP:conf/uss/YuanCZLL0ZH0G18}.

\subsection{Audio Adversarial Example Defense and Detection}

The emergence of adversarial examples has attracted researchers to study its defense strategies. Many studies on detecting image AEs have been reported, such as~\cite{zuo2018countermeasures}, while only a few are presented to cope with audio AEs, probably because techniques for crafting audio AEs just surfaced in the past two years. This also makes countermeasures against audio AEs urgent and important. 

%Researchers first try to leverage the input transformation as a defense technique, aiming to destroy the embedded hidden content. \cite{liang2018detect} proposed a defense method called \emph{feature squeezing} which treats the perturbations introduced by AE attacks as a kind of artificial noises and applies noise canceling techniques to reduce their effect. 
%Authors of the paper proved their method's effectiveness on image AEs and claimed it could be applied to any continuous data type including audios. 
%\cite{yangtowards} made a summary of input transformation-based defense schemes. 
Rajaratnam et al. \cite{rajaratnam2018AudioPreprocessing} proposed to detect audio AEs based on
audio pre-processing methods. Yet, if an attacker knows the detection details,
he can take the pre-processing effect into account when generating AEs. 
Such attacks have been well demonstrated for bypassing
similar techniques for detecting image AEs~\cite{bypassing-ten}.
\cite{athalye2018obfuscated} further pointed out that the input transformation only gives a false sense of robustness against AEs by imposing obfuscation over gradients which can be circumvented by their proposed method. 
%The circumvention was tested against ICLR 2018 non-certified defenses and the result showed those defense methods were still vulnerable. 
%Then, some works explore to 
%distinguish the AEs from benign audios. 
Carlini et al. \cite{DBLP:conf/uss/CarliniMVZSSWZ16} trained a logistic regression classifier, which was trained using a mix of benign and Hidden-Voice-Command (HVC) audios. But it can only detect hidden voice commands,
 instead of general audio AEs.

%But the classifier can only handle hidden voice commands
%instead of generate audio AEs~\cite{DBLP:conf/uss/CarliniMVZSSWZ16}. \cite{rajaratnam2018AudioPreprocessing} proposed a detector by combining various audio pre-processing methods. But an attacker can take the pre-processing effect into account when generating AEs; the technique has been well demonstrated in \cite{bypassing-ten} in the image domain. 
%This type of defense methods require many AE samples for training purpose which are usually not available as attackers won't publish the details of their attacks. Also, attackers can bypass these defense methods by simply changing some parameters or designing new attacking method. As a result, this type of defenses are weak in face of unknown attacks. 

Yang \textit{et al.} \cite{yang2018Mitigation} proposed to identify audio AEs based on the assumption that audio AEs need complete audio information to resolve temporal dependencies. They cut the input audio into two sections, which were then transcribed separately. If the input is an AE, the result obtained by splicing of the two sectional results will be very different from the result when the input is transcribed as a whole. However, as admitted by the authors, their method cannot handle ``adaptive attacks'', which evade the detection by only embedding malicious commands into one section alone. In short, the
literature has not reported an effective and robust audio AE detection method like ours.

%That is, transcribed texts of an benign audio and its first half are expected to have a very good match up to the length of . But, for an audio AE, they should differ significantly.

%This derived statement is supported by their experiment where audio AEs result into 95.8\% word error rate (WER) and 83.0\% character error rate (CER) on average, while, in comparison, benign audios show a largely reduced averaged WER (37.7\%) and CER (18.5\%). 
%This characteristic difference between benign and adversarial audios could potentially be used to find a decent decision border between them.

%investigate the defensive effect of various audio preprocessing methods (compression, speech coding, filtering and audio panning) and combine them to propose a detector that achieves a precision of 93.5\% and a recall of 91.2\%.

%To fight against the attack of black-box AEs, which are generated by a genetic algorithm and targets to mislead a widely adopted model, Speech Commands Model, Rajaratnam et al. \cite{rajaratnam2018AudioPreprocessing} investigate the defensive solution of applying various audio preprocessing methods and then derive an effective detector with a precision of 93.5\% and a recall of 91.2\%, which is an ensemble of compression, speech coding, filtering and audio panning.

\section{Discussion}\label{sec:future}
If the malicious command embedded in an AE and the host transcription are very similar, our method will probably fail as their similarity score is high. But note that, prior to our work, the existing AE generation methods claim that \emph{any} host audio can be used to embed a malicious command~\cite{DBLP:conf/sp/Carlini018,DBLP:journals/corr/abs-1805-07820}. Our detection method dramatically reduces this attack flexibility, in that the attack cannot succeed unless the host transcription is similar to the malicious command. %This makes audio AEs much less powerful than what was claimed.

The current prototype system successfully demonstrates the 
feasibility of the Multiversion Programming
inspired approach to detecting audio AEs, and the novel idea of 
proactively preparing a detection system resilient to transferable AEs, which
may be generated in future.
However, to deploy such a system still needs more engineering efforts.
For example, the online Amazon Transcribe ASR imposes large delays to return the transcription result
immediately, probably because of the server load control on the Amazon side.
However, we argue that such delays are inherent in our idea or system design. 
The delays as well as the deployment barrier can be eliminated 
by running multiple high-quality ASRs locally. For example, by
using \emph{DeepSpeech v0.1.0} and \emph{DeepSpeech v0.1.1}
to build the \textsc{MVP-Ears} system, it imposes negligible
delays (see Section~\ref{sec:overhead}) and achieves satisfactory accuracy
(99.56\%).

\section{Conclusion}\label{sec:con}
Research on handling audio AEs is still limited. 
Considering that ASRs are widely deployed in smart homes, smart phones and
cars, how to detect audio AEs is an important problem. 
Inspired by Multiversion Programming, we propose to run multiple different ASR systems in parallel,
and an audio input is determined as adversarial if the multiple ASRs generate very dissimilar transcriptions.
Detection systems with one single auxiliary ASR achieve satisfactory accuracies (\textgreater 98\%), while
systems with more than one ASR reach even higher accuracies (99.88\%) as more features are provided to
the classifier. 
The research results invalidate the widely-believed claim that an adversary can embed a malicious command to \emph{any} host audio. 

In addition, we propose the novel idea of training a proactive detection system for handling transferable audio AEs, such that the detection keeps effective as long as the AE is not able to
fool \emph{all} the ASRs in the detection system. Therefore,
it makes our system one stride ahead of attackers working on generating transferable audio AEs.

%To defend the-state-of-art attack of targeted audio AEs, we proposed a system design that utilizes off-the-shelf ASR(s) as auxiliary model(s) to detect an incoming AE. Several experiments are conducted upon a dataset with 1125 benign and 1125 adversarial audios. DeepSpeech v0.1.1, Google Cloud Speech and Amazon Transcribe are picked as auxiliary models to protect the target model, DeepSpeech v0.1.0. PE\_JaroWinkler stands out from a comparison experiment and is selected as our similarity metric. In experiments, all single-auxiliary-model systems achieve an accuracy of higher than 98.6\% with both FPR and FNR of lower than 1.34\%, while Multiple-Auxiliary-Models systems demonstrate better performance in general. The Three-Auxiliary-Models system with either KNN or Random Forest as the classifier reaches an accuracy of 100\%. Besides, the experiment with unknown AEs further justifies the robustness of the proposed system design.

%\input{untargeted.tex}

\section*{Acknowledgment}
This project was supported by NSF CNS-1815144 and NSF CNS-1856380.
The authors would like to thank Nicholas Carlini for sharing their AEs~\cite{DBLP:conf/sp/Carlini018},
and anonymous reviewers for their comments and suggestions.
In the early stage of this project, we used the Chameleon Cloud (funded by NSF),
so we would like to thank this project~\cite{chameleon}.

\balance
%References and End of Paper
%These lines must be placed at the end of your paper
\bibliographystyle{IEEEtran}
\bibliography{IEEEabrv,references} 

\end{document}